\documentclass[useAMS,usenatbib]{mn2e}

\usepackage{graphicx}
\usepackage{amssymb}
\usepackage{multirow}

\makeatletter
    
    \newcommand{\Rmnum}[1]{\expandafter\@slowromancap\romannumeral #1@}
\makeatother

\title[A study of the high-inclination population in the Kuiper belt -- \Rmnum2. The Twotinos]
         {A study of the high-inclination population in the Kuiper belt -- \Rmnum2. The Twotinos}
\author[Jian Li, Li-Yong Zhou and Yi-Sui Sun]
{Jian Li\thanks{E-mail: ljian@nju.edu.cn},  Li-Yong Zhou and Yi-Sui Sun\\
School of Astronomy and Space Science
\& Key Laboratory of Modern Astronomy and Astrophysics in Ministry of Education,\\
 Nanjing University, Nanjing 210093, PR China}
\begin{document}

\date{Accepted 1988 December 15. Received 1988 December 14; in original form 1988 October 11}

\pagerange{\pageref{firstpage}--\pageref{lastpage}} \pubyear{2002}

\maketitle

\label{firstpage}

\begin{abstract}

As the second part of our study, in this paper we proceed to explore the dynamics of the high-inclination Twotinos in the 1:2 Neptune mean motion resonance (NMMR). Depending on the inclination $i$, we show the existence of two critical eccentricities $e_a(i)$ and $e_c(i)$, which are lower limits of the eccentricity $e$ for the resonant angle $\sigma$ to exhibit libration and asymmetric libration, respectively. Accordingly, we have determined the libration centres $\sigma_0$ for inclined orbits, which are strongly dependent on $i$. With initial $\sigma=\sigma_0$ on a fine grid of $(e, i)$, the stability of orbits in the 1:2 NMMR is probed by 4-Gyr integrations. It is shown that symmetric librators are totally unstable for $i\ge30^{\circ}$; while stable asymmetric librators exist for $i$ up to $90^{\circ}$. 

We further investigate the 1:2 NMMR capture and retention of planetesimals with initial inclinations $i_0\le90^{\circ}$ in the planet migration model using a time-scale of $2\times10^7$ yr. We find that: (1) the capture efficiency of the 1:2 NMMR decreases drastically with the increase of $i_0$, and it goes to 0 when $i_0\gtrsim60^{\circ}$; (2) the probability of discovering Twotinos with $i>25^{\circ}$, beyond observed values, is roughly estimated to be $\lesssim0.1$ per cent; (3) more particles are captured into the leading rather than the trailing asymmetric resonance for $i_0\le10^{\circ}$, but this number difference appears to be the opposite at $i_0=20^{\circ}$ and is continuously varying for even larger $i_0$; (4) captured Twotinos residing in the trailing resonance or having $i>15^{\circ}$ are practically outside the Kozai mechanism, like currently observed samples.  

 \end{abstract}


%_____________________________________________________________________________________________________________________

\section{Introduction}

The Kuiper belt objects (KBOs) are icy celestial bodies near and beyond the orbit of Neptune in the outer Solar system. One of the most outstanding mysteries of these distant objects is the unexpected high orbital inclination ($i$). In the previous work (Li et al. 2014, hereafter LZS14), we began to study this issue by considering the dynamics of the high-inclination Plutinos. By definition, a Plutino occupies the 2:3 Neptune mean motion resonance (NMMR) at semimajor axis $a\approx39.4$ au. We have defined the special libration centre (SLC) of the resonant angle $\sigma_{2:3}$ at the minimum of the averaged disturbing function $R({\sigma_{2:3}})$, which corresponds to a stable equilibrium point because the variation of the semimajor axis $da/dt(\propto{\partial R}/{\partial \sigma_{2:3}})=0$. Then we designated the mean value of $\sigma_{2:3}$ during the time evolution as the general libration centre (GLC) at $180^{\circ}$, i.e., the usually called ``libration centre''. For low-inclination orbits, the position of the SLC is fixed exactly at the GLC. But for high-inclination orbits, the SLC becomes strongly dependent on the argument of perihelion $\omega$ and oscillates around the GLC with large amplitude (Gallardo 2006). Nevertheless, the average value of the SLC over $0^{\circ}\le \omega \le360^{\circ}$ always matches the GLC. This outcome suggests a new method that can help us to easily determine the libration centre of the other NMMRs for any inclined orbits. We refer the reader to LZS14 for more details on the calculations of the SLC.

In LZS14 we also updated the possible $i$-range of the potential Plutinos.  Using the $N$-body simulations, we first investigated the stability of high-inclination candidates populating in the 2:3 NMMR for times up to the age of the Solar system. The dynamical map has been built up on the initial $(a, i)$ plane at different eccentricities ($e$). We have found that the stable resonant orbits could cover the whole inclination space of $i\le90^{\circ}$. Based on this result, we further explored the outward transportation of planetesimals with initial $i$ up to $90^{\circ}$ in the framework of the planet migration and 2:3 NMMR sweeping model (Malhotra 1995). Our results showed that the resonant capture and retainment is allowed for any inclined or even perpendicular orbits. Beside the aforementioned main concerns, the role of high $i$ in both the formation of Plutinos during Neptune's migration and the later long-term evolution has been discussed in detail.

\begin{figure}
 \hspace{-0.7cm}
  \centering
  \includegraphics[width=9cm]{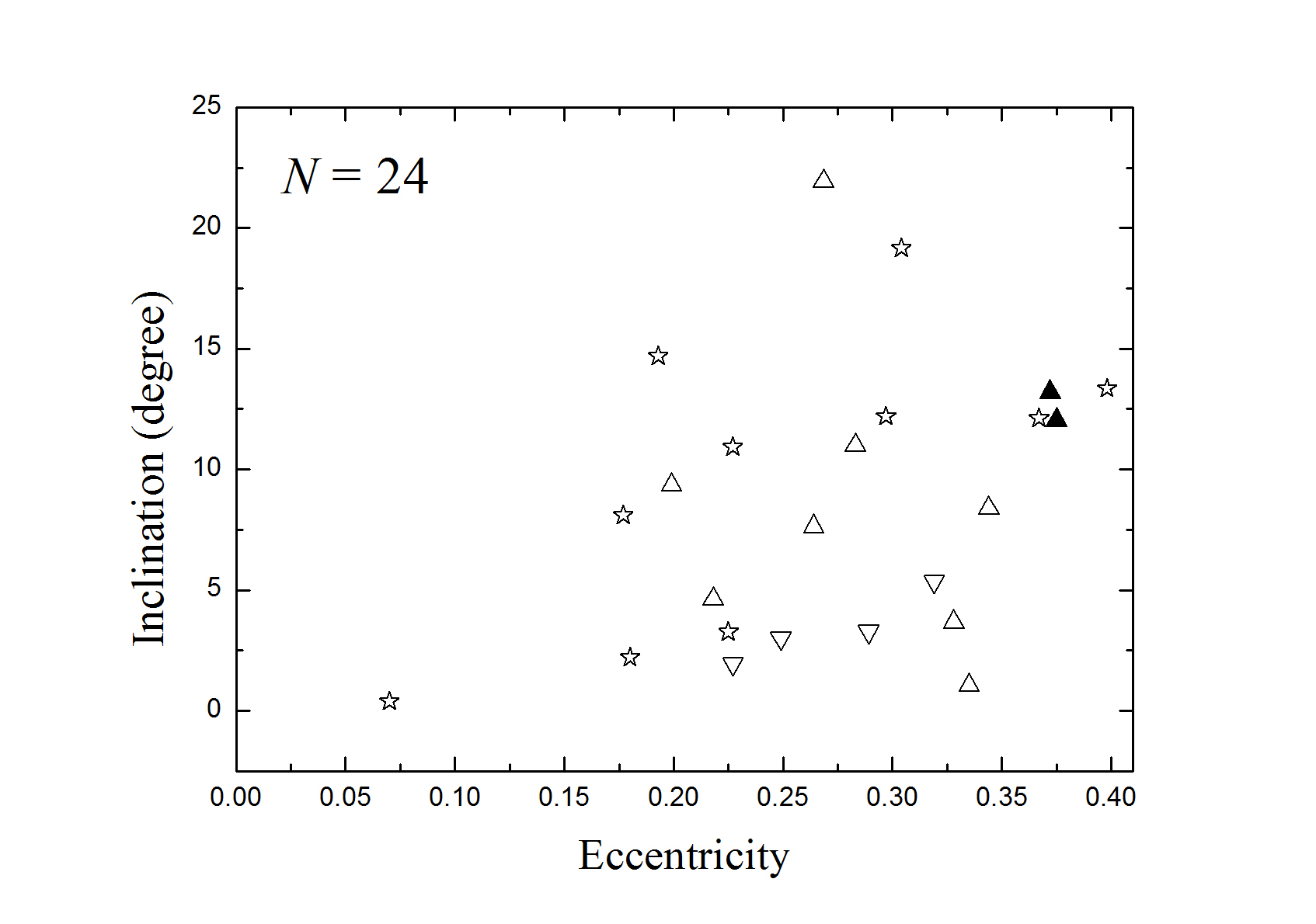}
  \caption{Distribution of eccentricities and inclinations for the currently observed Twotinos (as of October 2012). The symmetric librators are stars, the leading librators are upward triangles, and the trailing librators are downward triangles. The filled symbols refer to Twotinos experiencing the Kozai mechanism (see Section 3.2.2 for definition).}
  \label{observed}
\end{figure}

As the continuation of our study of the high-inclination KBOs, in this paper we investigate a population of bodies with $a\approx47.8$ au in the 1:2 NMMR, which are known as Twotinos. It is of further interest to present a global view of the dynamics of Twotinos on inclined orbits. In the MPC (Minor Planet Center) database\footnote{http://www.minorplanetcenter.net/iau/lists/TNOs.html}, about 24 Twotinos have been registered and their osculating $e$ and $i$ at epoch 2012 September 30 are shown in Fig. \ref{observed}. For the sake of identifying these Twotinos, we numerically integrated the trajectories of observed KBOs with $a=47.8\pm0.8$ au for a timespan of $10^8$ yr, under the gravitational perturbations of four Jovian planets. Then we examined the time evolution of the resonant angle $\sigma=\lambda_N-2\lambda+\varpi$ corresponding to the 1:2 NMMR, where $\lambda$ and $\varpi$ are the mean longitude and the longitude of perihelion of the KBO, respectively, and $\lambda_N$ is the mean longitude of Neptune. In this way a KBO is regarded as a Twotino if it exhibits libration of $\sigma$, i.e., the resonant amplitude $A_{\sigma}<180^{\circ}$. It is seen that the Twotinos possess high inclinations up to about $25^{\circ}$, which seems to be in accord with the stable $i$-range obtained in previous literatures (Melita \& Brunini 2000; Nesvorn\'{y} \& Roig 2001; Lykawka \& Mukai 2007; Tiscareno \& Malhotra 2009; Gladman et al. 2012). In spite of this, we are still eager to know whether $i\sim25^{\circ}$ is actually the limit of all the potential Twotinos. It must be noted that here we just confirmed or rejected the 1:2 NMMR librators with nominal orbital elements given in the MPC site. If one takes into account the uncertainties of the orbital elements, there might yield some other librators near 47.8 au, and particularly a part of them could have $i>25^{\circ}$. Besides, observational selection effects would also work against the detection of high-inclination objects.

It is well known that, for the type of 1:$n$ exterior mean motion resonances, there are three possible resonant modes: the usual {\it symmetric} libration of $\sigma$ around $180^{\circ}$; and two separate {\it asymmetric} librations for $0^{\circ}<\sigma<180^{\circ}$ (leading) and for $180^{\circ}<\sigma<360^{\circ}$ (trailing) (Beaug\'{e} 1994; Morbidelli et al. 1995).  The leading and trailing librators reach their perihelia at longitudes ahead of and behind Neptune's longitude, respectively. In the framework of the planar circular restricted 3-body problem, Malhotra (1996) pointed out that the libration centre ${\sigma}_0$ is not a well-defined one but depends upon the orbital eccentricity $e$ of the particle. Later, the value of ${\sigma}_0$ related to the asymmetric 1:2 NMMR as a function of $e$ was estimated by Nesvorn\'{y} \& Roig (2001). However, the dependence of ${\sigma}_0$ on the inclinations of Twotinos has not been researched until now. This is due to the fact that the analytical or semi-analytical computations of asymmetric periodic orbits are not so easy in the three-dimensional restricted 3-body problem (Kotoulas 2005). 

With respect to the history of Neptune's outward migration, the relative populations of the leading and trailing Twotinos may be an important diagnostic for the migration time-scale $\tau$. Chiang \& Jordan (2002) and Murray-Clay \& Chiang (2005) proposed that planetesimals are less likely to be trapped into the leading rather than the trailing 1:2 NMMR for $\tau$ shorter than $\sim10^7$ yr. But according to observational data shown in Fig. \ref{observed}, 10 Twotinos are in the leading resonance and only 4 are in the trailing resonance, such a number difference may support an even larger $\tau$ (Lykawka \& Mukai 2007). Indeed, a time-scale of $\tau=2\times10^7$ yr was derived by Li et al. (2011) from $N$-body simulations for the orbital evolution of Jovian planets embedded in a self-gravitating planetesimal disc. This reasonable value has already been adopted in LZS14. 

Moreover, Murray-Clay \& Chiang (2005) also discussed the capture into the asymmetric 1:2 NMMR influenced by planetesimal's initial eccentricity $e_0$ and initial semimajor axis $a_0$. Given the time-scale of $\tau=10^7$ yr and $a_0\ge37.7$ au (1 au exterior to the initial 1:2 NMMR), the ratio of leading to trailing particles can be infinitely small for $e_0=0.01$; while it grows with lager $e_0$ and would nearly approach 1 for $e_0=0.05$, i.e., the difference in populations vanishes. As we argued in LZS14, the primordial planetesimals may have high initial inclinations $i_0$ prior to the onset of the resonance sweeping. Therefore, in order to complete a survey of the 1:2 NMMR capture in the entire orbital element space, a detailed analysis of the role of $i_0$ is clearly warranted.

The rest of this paper is organized as follows. In Section 2, we reveal the shift of the libration centre ${\sigma}_0$ due to high $i$, and the complex resonant behavior in the 1:2 NMMR. In Section 3, we probe the long-term stability of high-inclination objects in the 1:2 NMMR, and show that $i$ of stable librators could be as high as $90^{\circ}$. Here, the $i$-dependent ${\sigma}_0$ is crucial to choose the initial conditions. In Section 4, for different initial inclinations $i_0$, we discuss in detail the resonance capture and orbital evolution of planetesimals in the planet migration model using a time-scale of $\tau=2\times10^7$ yr. In Section 5, we conclude this paper with a summary of our main findings, and discuss the theoretical models and current observations.

%_____________________________________________________________________________________________________________________

\section[]{The resonant behavior and the libration centre}

\begin{figure}
 \hspace{-0.7cm}
  \centering
  \includegraphics[width=9cm]{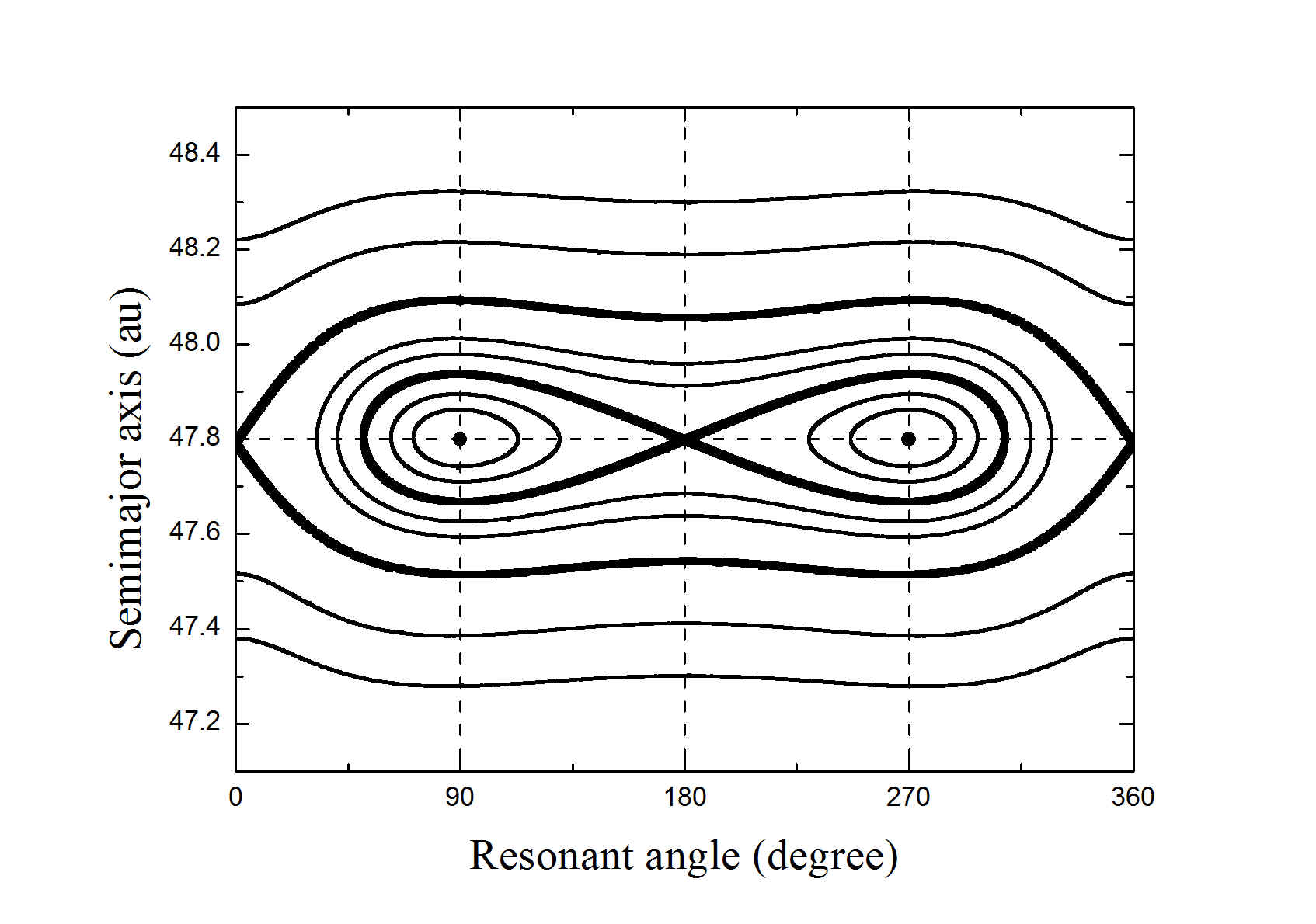}
  \caption{The phase space of the 1:2 NMMR for eccentricity $e=0.15>e_c$ in the framework of the planar circular restricted 3-body problem. The horizontal dashed line at 47.8 au denotes the location of nominal 1:2 NMMR. The three vertical dashed lines indicate the symmetric libration centre at $180^{\circ}$, and the asymmetric ones at $90^{\circ}$ (leading) and $270^{\circ}$ (trailing).}
  \label{phase}
\end{figure}

Restricted to the planar case ($i=0^{\circ}$), the dependence of the libration centres of the 1:2 NMMR on the eccentricity $e$ has been published for many years and is known perfectly well (Beaug\'{e} 1994; Morbidelli et al. 1995; Malhotra 1996). At small values of $e$, the phase space of the 1:2 NMMR is largely regular like a pendulum system, and there is only a symmetric libration centre $\sigma^{\pi}_0$ at $180^{\circ}$. While for $e$ exceeding a critical value $e_c\sim 0.04$, the typical $(\sigma, a)$ phase space of this resonance is sketched in Fig. \ref{phase}. The symmetric resonance island splits into two asymmetric resonance islands, where the resonant amplitudes $A_{\sigma}$ have an upper bound ($<90^{\circ}$). Meanwhile, the symmetrically librating orbits with larger $A_{\sigma}$ (``horseshoe'' orbits) can still exist, surrounding the asymmetrically librating orbits (``tadpole'' orbits). Note that the locations of asymmetric centres can be modified by the value of $e$. For $e=0.15>e_c$, as shown in Fig. \ref{phase}, the leading centre $\sigma^L_0$ and the trailing centre $\sigma^T_0$ are placed at $90^{\circ}$ and $270^{\circ}$, respectively.

\begin{figure}
  \centering
  \begin{minipage}[c]{0.6\textwidth}
  \centering
  \hspace{-3cm}
  \includegraphics[width=9cm]{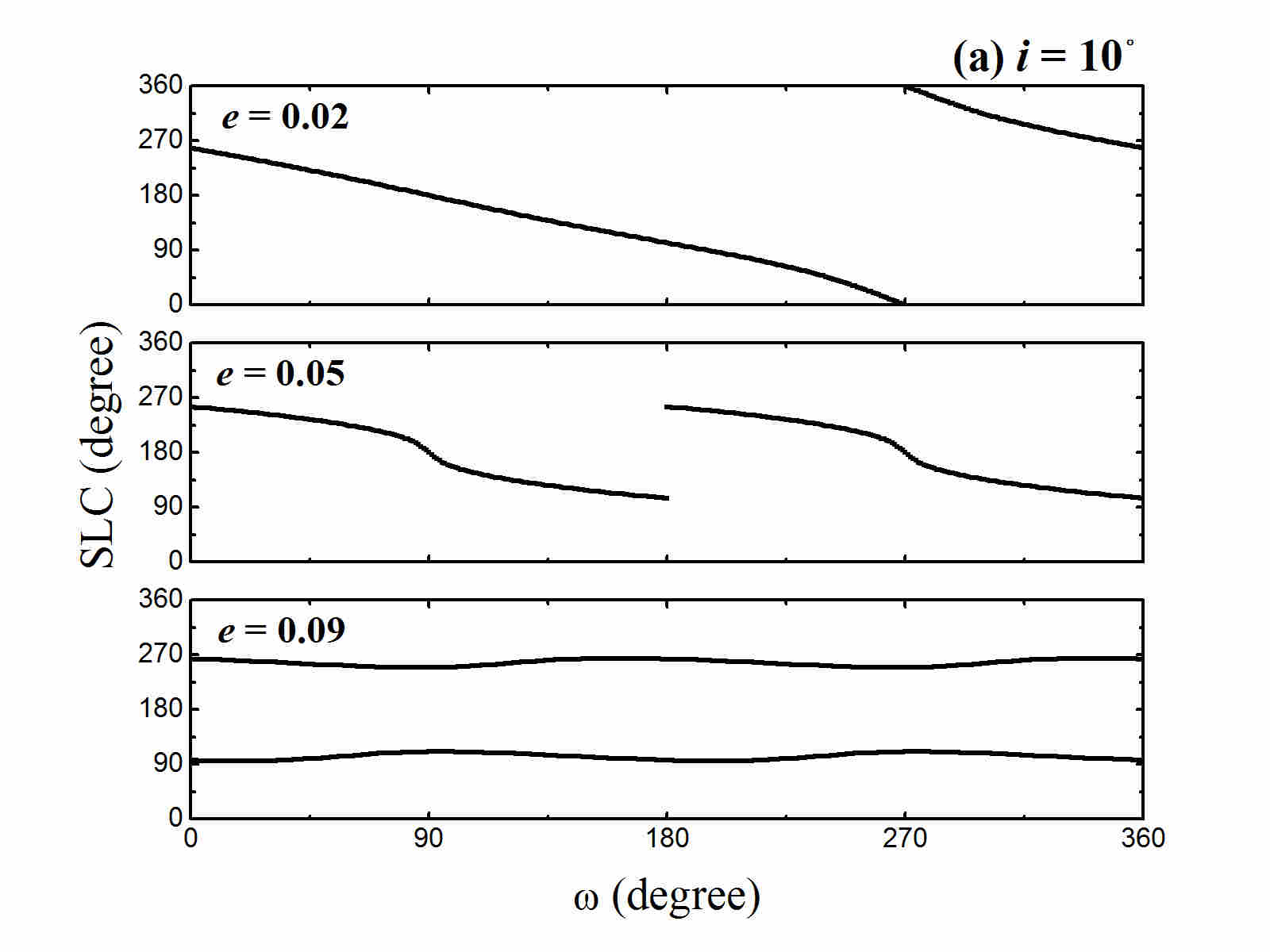}
  \end{minipage}
  \begin{minipage}[c]{0.6\textwidth}
  \centering
  \hspace{-3cm}
  \includegraphics[width=9cm]{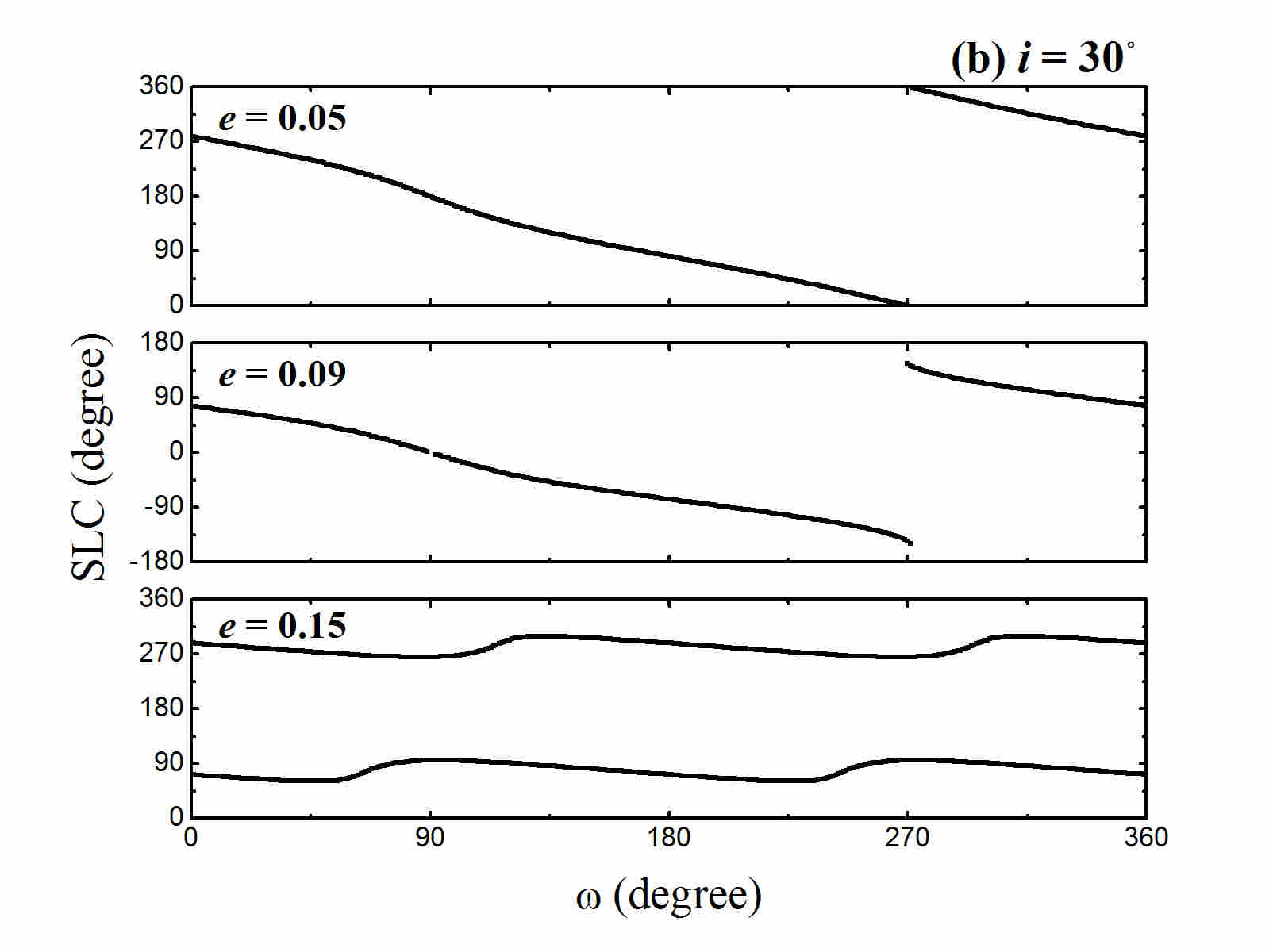}
  \end{minipage}
  \caption{The behavior of the SLC for the 1:2 NMMR for (a) $i=10^{\circ}$ and (b) $i=30^{\circ}$. The upper, middle and lower panels correspond to the cases of $e<e_a$, $e_a<e<e_c$ and $e>e_c$, respectively. In the middle panel of graph (b), notice at $\omega=270^{\circ}$ that the SLC is approximately $34 ^{\circ} $ away from $\pm180^{\circ}$.}
 \label{SLC}
\end{figure}

To evaluate the libration centres of the 1:2 NMMR for inclined orbits, we followed the semi-analytical method developed in LZS14. The libration centre is denoted by the GLC, i.e., ${\sigma}_0=({\sigma_{max}+{\sigma}_{min}})/2$, which can be simply and precisely calculated as the mean value of the SLC. And for the resonant amplitude $A_{\sigma}=\sigma_{max}-{\sigma}_0$, it has a lower limit that is virtually the amplitude of the SLC. Unlike in the planar case, for a specific $e$, the SLC would be no longer fixed at the GLC but moves right along with the argument of perihelion $\omega$ when the non-zero $i$ has been introduced (Gallardo 2006).

The general characteristics of the 1:2 NMMR can be understood by analyzing the behavior of the SLC for a series of $\omega$ between $0^{\circ}$ and $360^{\circ}$, as presented in Fig. \ref{SLC}:

\begin{enumerate}
  \item If $e<e_a$, the motion of the SLC is unbounded (Fig. \ref{SLC}, upper panels). Since the amplitude of the SLC is the lower limit of $A_{\sigma}$, this implies that the resonant angle $\sigma$ would eventually take all values between $0^{\circ}$ and $360^{\circ}$.  As we will see in the next section, the resonant angle $\sigma$ would be simply circulating in this case; while a plausible scenario where there is a libration of $\sigma$ around SLC but SLC itself circulates would never occur. Hence we consider the critical eccentricity $e_a$ to correspond to the lower limit of $e$ for librations in the 1:2 NMMR.
  
  \item If $e_a<e<e_c$, then the motion of the SLC becomes bounded. For relatively small $i$ (e.g., Fig. \ref{SLC}a, middle panel), the SLC does not pass by the location of $0^{\circ}$, and shows libration around the GLC at $180^{\circ}$. But for relatively large $i$,  it is found that the SLC never achieves $180^{\circ}$ and then the GLC changes to $0^{\circ}$. This is illustrated by the middle panel of Fig. \ref{SLC}b, where the interval of the vertical axis is adopted to be [-$180^{\circ}$, $180^{\circ}$] for a simpler visualization. Based on our calculations, the shift of the GLC from $180^{\circ}$ to $0^{\circ}$ takes place at a tentative inclination of $\sim15^{\circ}$. Note that the resonant orbits with small $A_{\sigma}$ ($<90^{\circ}$) are permitted here.\\
For the perihelic conjunctions at $\sigma=0^{\circ}$, the vertical distance from a small-$A_{\sigma}$ Twotino to Neptune's orbital plane is roughly estimated by $d=a(1-e_c)\sin{i}$. This yields a value of $d\approx11.4$ au for $i=15^{\circ}$, and this distance is monotonically increasing with $i$. Therefore, in the absence of strong gravitational perturbation from Neptune, the libration of $\sigma$ around $0^{\circ}$ could be a possibly stable configuration.
  
  \item If $e>e_c$, the motion of the SLC is restricted to either the interval of $[\sigma_1, \sigma_2]\subset (0^{\circ}, 180^{\circ})$  or $[\sigma_3, \sigma_4] \subset (180^{\circ}, 360^{\circ})$ (Fig. \ref{SLC}, lower panels). This implies the appearance of asymmetric islands in the 1:2 NMMR (see Fig. \ref{phase}), and there are two GLCs: $\sigma^L_0=(\sigma_1+\sigma_2)/2$ and $\sigma^T_0=(\sigma_3+\sigma_4)/2$. As mentioned earlier, $e_c$ represents the low limit for $e$ for the permission of the asymmetric 1:2 NMMR. It is necessary to stress that, our SLCs are determined at the minimum of the resonant disturbing function, but they do not exclude the existence of large-$A_{\sigma}$ symmetric librations around the GLC $\sigma^{\pi}_0=180^{\circ}$ (Gallardo 2006).
\end{enumerate}

 \begin{figure}
 \hspace{-0.7cm}
  \centering
  \includegraphics[width=9cm]{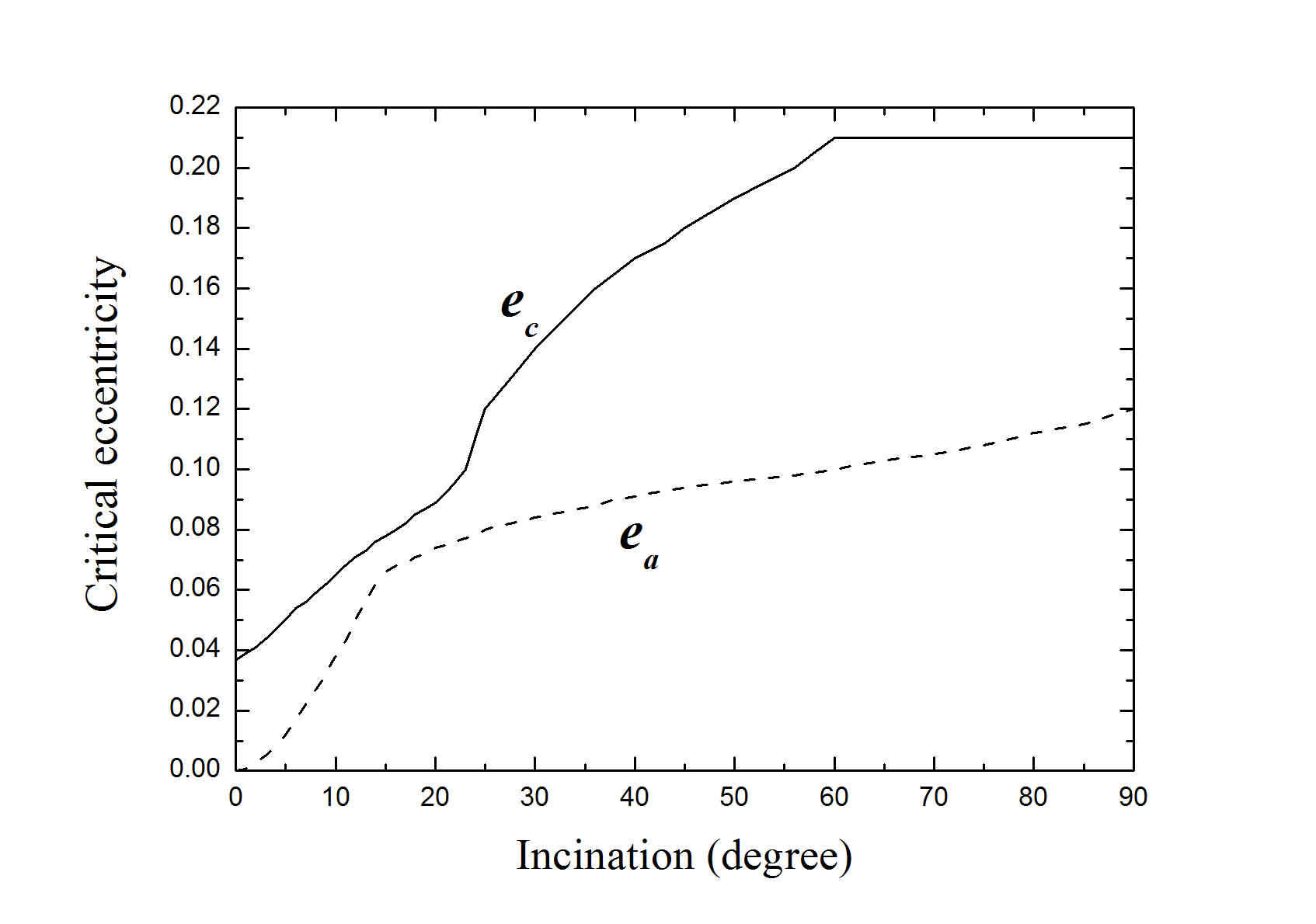}
  \caption{The critical eccentricities $e_a$ (dashed line) and $e_c$ (solid line) for the 1:2 NMMR as functions of the inclination $i$. Note that these two quantities both increase with $i$.}
  \label{eaec}
\end{figure}

The two critical eccentricities $e_a$ and $e_c$ for the 1:2 NMMR are both functions of the inclination $i$. Fig. \ref{eaec} gives the values of $e_a(i)$ and $e_c(i)$ derived from the different behaviors of the SLCs for orbits with $i$ from $0^{\circ}$ to $90^{\circ}$. For the planar case of $i=0^{\circ}$, we have $e_a=0$ and $e_c=0.037$, which are in reasonably good agreement with those obtained by Beaug\'{e} (1994) and Malhotra (1996). With increasing $i$, these two quantities both shift to larger values, and $e_c$ reaches its peak of $\sim0.21$ when $i$ exceeds about $60^{\circ}$. Especially, for the highest inclination of $\sim25^{\circ}$ observed in Fig. \ref{observed}, the corresponding $e_c$ is about 0.12.  This may agree with the fact that real Twotinos preferentially have large eccentricities. The only exception is a Twotino with $e\sim0.07$ in the bottom-left corner of Fig. \ref{observed}. Due to the nearly zero inclination, it has sufficiently large $e$ relative to $e_c(i=0^{\circ})=0.037$. 

 \begin{figure}
 \hspace{-0.7cm}
  \centering
  \includegraphics[width=9cm]{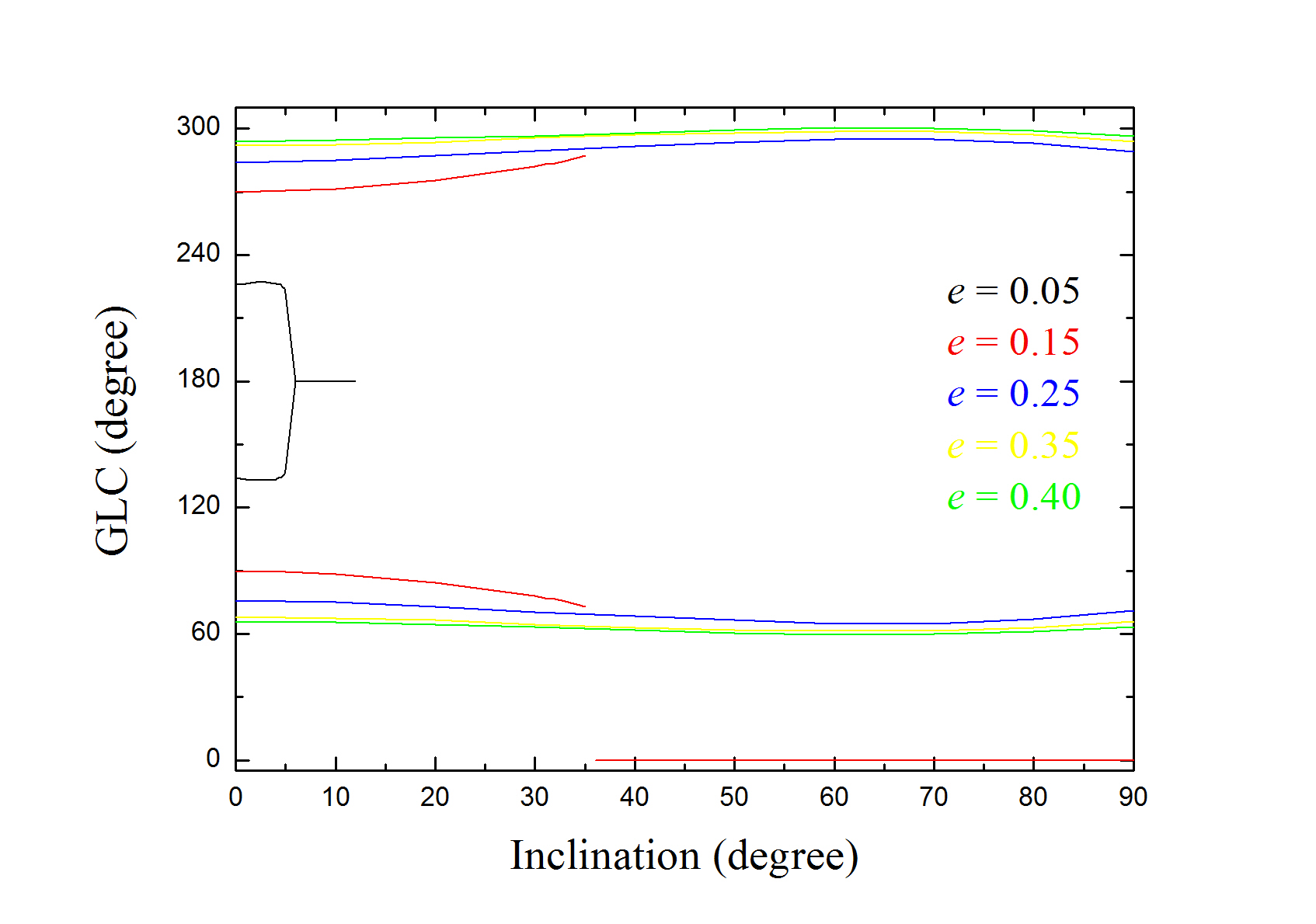}
  \caption{The locations of the GLC with respect to inclination for the 1:2 NMMR, at five eccentricities of $e=0.05$ (black), 0.15 (red), 0.25 (blue), 0.35 (yellow) and 0.40 (green).}
  \label{GLC}
\end{figure}

Taking advantage of the above calculations, we present in Fig. \ref{GLC} the locations of the GLC with respect to inclination for the 1:2 NMMR, at five representative eccentricities of $e=0.05$, 0.15, 0.25, 0.35 and 0.40 (covering the observational range). The resonant states of Twotinos can be divided into the following three typical categories:

\begin{enumerate}
  \item For $e=0.05$, the asymmetric libration has ceased to exist when $i>5^{\circ}$, while only the symmetric libration around $180^{\circ}$ is possible. Furthermore, if $i$ continues to grow beyond about $12^{\circ}$, objects would be surely outside the 1:2 NMMR since the libration of $\sigma$ is prohibited by the circulation of the SLC.
  
  \item For $e=0.15$, the asymmetric libration is allowed when $i\le35^{\circ}$. We find that the libration centre $\sigma^L_0$ ($\sigma^T_0$) is at $90^{\circ}$ ($270^{\circ}$) for $i=0^{\circ}$, exactly as that shown in Fig. \ref{phase}, and it decreases (increases) continuously to $73^{\circ}$ ($287^{\circ}$) as $i$ increases to $35^{\circ}$. But for even larger $i$, equivalent to the case of $e_a<e<e_c$, it appears an uncommon libration centre at  $0^{\circ}$, which will be further discussed in a later section.
  
  \item For $e\ge0.25$, the asymmetric libration is possible for any value of $i$, because $e$ is larger than the maximum $e_c(i)\sim0.21$. Along with the increase of $i$, the magnitude of variation in $\sigma^L_0$ and $\sigma^T_0$ is either above or near $10^{\circ}$. 

\end{enumerate}

For the numerical investigation of the long-term stability of high-inclination Twotinos, our semi-analytical predictions could provide a wealth of information on the choice of initial conditions. Accordingly, for a given $e$, we can set the initial resonant angle $\sigma$ of all orbits to be the exact GLC value within a proper $i$-range.

%_____________________________________________________________________________________________________________________

\section{The stability of the overall inclinations}

\subsection{Initial conditions and pre-runs}

\begin{table}
\centering
\begin{minipage}{8cm}
\caption{The initial conditions of the eccentricity $e$, the inclination $i$ and the resonant angle $\sigma$ for test particles. The values of the asymmetric libration centres $\sigma^L_0$ and $\sigma^T_0$ are directly drawn from Fig. \ref{GLC}. The last column refers to the resonant states of particles near 47.8 au in the pre-runs (see text for a definition of these letters).}      % title of Table
\label{initial}
\begin{tabular}{c c c c c}        % centered columns (9 columns)
\hline                 % inserts double horizontal lines
Case     &         $e$      &        $i(^{\circ})$             &                                                    $\sigma$                                                             &             State                      \\

\hline

    1         &         0.05     &                 0                       &                                  $\sigma^L_0$, $\sigma^T_0$                                          &       S\&A,  S\&A                \\
    
    2         &         0.05     &                 0                       &                                               $180^{\circ}$                                                          &                  S                            \\
        
    3         &         0.05     &                10                      &                                               $180^{\circ}$                                                          &                  S                             \\
            
    4         &         0.05     &             20--90                  &                                                $180^{\circ}$                                                         &                 C                              \\
    
\hline
    
    5         &         0.15     &               0--30                   &                     $\sigma^L_0$, $\sigma^T_0$, $180^{\circ}$                           &               S, A, A                         \\
          
    6         &         0.15     &              40--90                  &                                                $0^{\circ}$                                                              &                   C                               \\
    
\hline
    
    7         &         0.25     &   \multirow{3}{*}{0--90}  &         \multirow{3}{*}{ $\sigma^L_0$, $\sigma^T_0$, $180^{\circ}$}        &    \multirow{3}{*}{S, A, A}       \\
    
    8         &         0.35     &                 {}                         &                                     	               {}                                                                     &                    {}                                \\
     
    9         &          0.4      &                  {}                        &       			                         {}                                                                     &                    {}                                 \\

\hline
\end{tabular}
\end{minipage}
\end{table}

The initial conditions of the eccentricity $e$, the inclination $i$ and the resonant angle $\sigma$ for test particles are based on what we have just learned in the previous section (e.g., Fig. \ref{GLC}), and they are listed in Table \ref{initial}. Bearing in mind that the symmetric libration around $180^{\circ}$ can exist after the appearance of the asymmetric libration, we sampled particles starting at $\sigma=180^{\circ}$ together with $\sigma^L_0$ and $\sigma^T_0$ for $e>e_c(i)$. However, these orbital parameters are chosen to satisfy the resonance condition in the restricted 3-body problem, where the perturbation of other Jovian planets besides Neptune is not taken into account. 

In order to test whether the initial conditions in Table \ref{initial} can result in resonant orbits, we conducted a series of pre-runs under the Solar system model described in LZS14. For each ($e, i, \sigma$) set, 321 test particles are uniformly spaced between $47.8-0.8$ and $47.8+0.8$ au, with the same $\omega$ and longitude of ascending node ($\Omega$) as Neptune's. For cases 4--9, the increment in $i$ is adopted to be $10^{\circ}$. Using the SWIFT\_RMVS3 symplectic integrator (Levison \& Duncan 1994) with a time-step of 0.5 yr, the particles' orbits are integrated up to $10^6$ yr, which is at least 10 times longer than the typical libration period of $\sigma$ (Malhotra 1996). 

Corresponding to each set of initial ($e, i, \sigma$), the behavior of $\sigma$ near the location of the nominal 1:2 NMMR at $\sim47.8$ au has been illustrated in the last column of Table \ref{initial}. The letters ``S'', ``A'' and ``C'' mark symmetric libration, asymmetric libration and circulation, respectively; and the combination ``S\&A'' indicates the alternation between symmetric and asymmetric resonances. On comparing with the anticipated resonant states at the end of Section 2, one can immediately visualize the predictive validity of our semi-analytical approach. We want to emphasize that, here and hereafter, the C state refers to a real circulation of $\sigma$ in our numerical simulations, but not a libration of $\sigma$ with a circulating SLC in the libration timescale.

\begin{figure*}
  \centering
  \vspace{0.5 cm}
  \begin{minipage}[c]{0.9\textwidth}
  \centering
  \includegraphics[width=20cm]{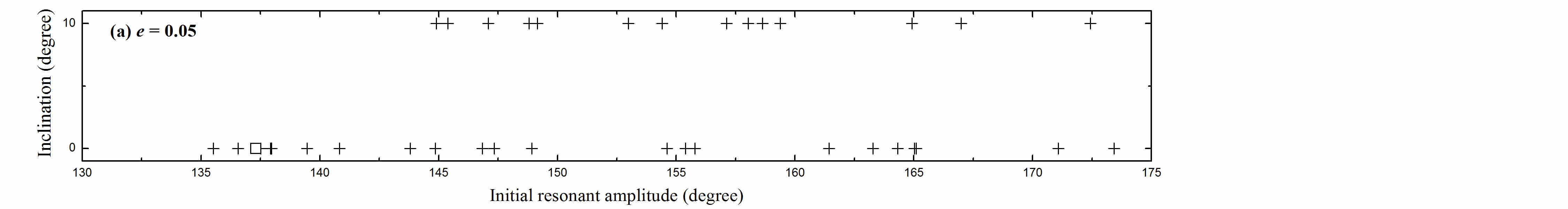}
  \end{minipage}
  \begin{minipage}[c]{0.9\textwidth}
  \centering
  \includegraphics[width=20cm]{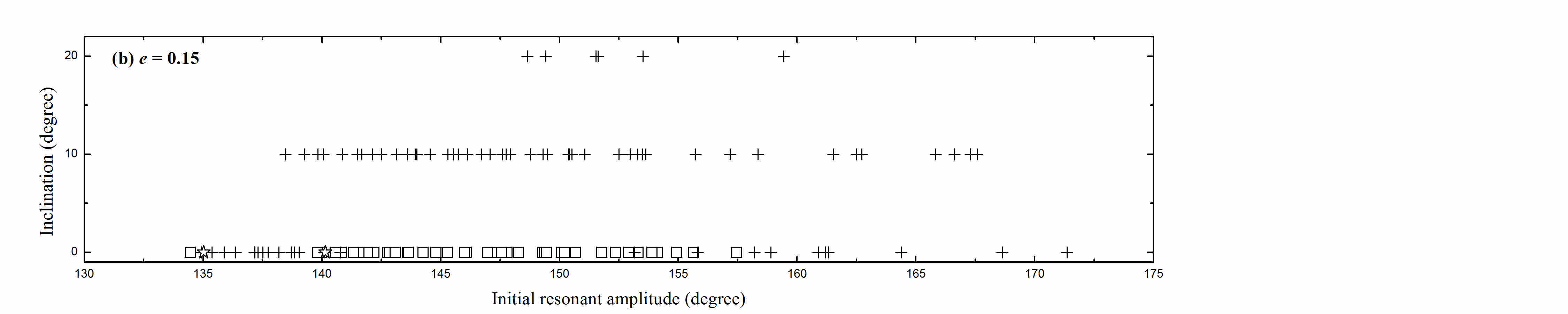}
  \end{minipage}
  \begin{minipage}[c]{0.9\textwidth}
  \centering
  \includegraphics[width=20cm]{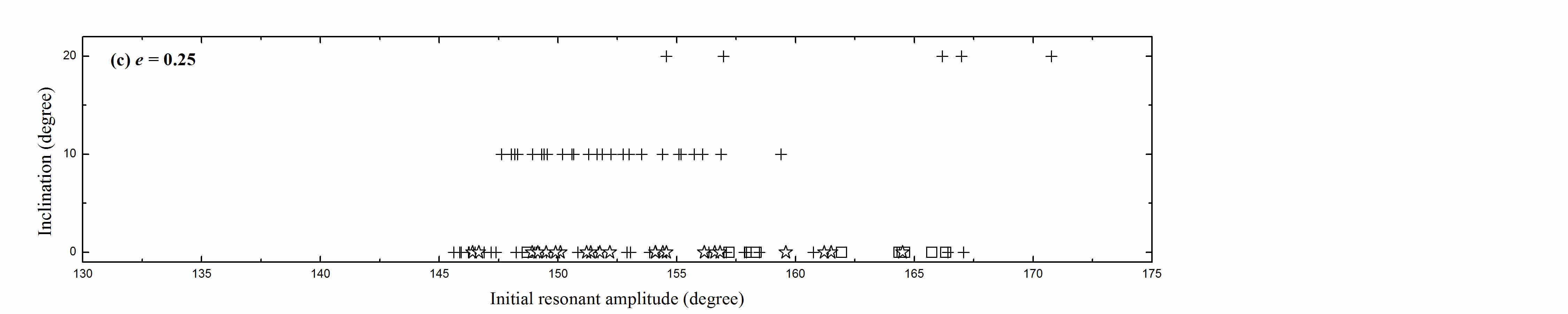}
  \end{minipage}
  \caption{The resonant states for initial symmetric librators from the pre-run in the $4\times10^9$ yr numerical simulations: (a) $e=0.05$; (b) $e=0.15$; (c) $e=0.25$. Squares represent particles in the persistent symmetric resonance, stars denote the alternation between symmetric and asymmetric librations, and crosses denote the transition between libration and circulation.}
 \label{sym}
\end{figure*}  

In Table \ref{initial}, we also notice two discrepancies that have to be addressed. The first is the case 1 for test particles with $e=0.05$ and $i=0^{\circ}$. A careful checking on the time evolution of $\sigma$ shows that, these particles originated from asymmetric orbits would evolve into S\&A orbits. In reality, Nesvorn\'{y} \& Roig (2001) argued that asymmetric motion in the 1:2 NMMR could not maintain in the range $e<0.1$ at low $i$, while symmetric motion with such small $e$ may still be stable on time-scales of several Gyrs. Second, in the case 6 of initial $e=0.15$ and $i\ge40^{\circ}$, all particles are found to end up on circulation rather than libration around $0^{\circ}$. This inconsistency is essentially due to the combined gravitational effect of all Jovian planets. As an illustrative example, we performed additional runs within the Sun+Neptune+particle restricted 3-body model. Beginning with exactly the same initial conditions as in case 6, test particles librating around $\sigma=0^{\circ}$ can survive in the $10^6$ yr integration, and some of them have small $A_{\sigma}$. However, if we just change the initial resonant angle $\sigma$ to $180^{\circ}$, all particles would display persistent circulation.

As a consequence, we will restrict ourselves to consider only the stably librating particles with $A_{\sigma}<180^{\circ}$ in the cases 2, 3, 5 and 7--9 for the long-term run below. Moreover, in order to discuss the dynamical stabilities of tadpole and horseshoe orbits separately, our choice of the initial $a$ must cover the 1:2 NMMR zone properly. Fig. \ref{phase} shows that, if $\sigma$ is chosen at the libration centre, then $a$ is related to the resonant amplitude $A_{\sigma}$. In particular, test particles started at $\sigma^L_0$ or $\sigma^T_0$ may actually happen on horseshoe orbits, provided they are relatively far from 47.8 au. At this point, we narrow the $a$-width of the 1:2 NMMR to include only the asymmetric librators with $A_{\sigma}<90^{\circ}$. 

After recording for each librating particle the resonant amplitude $A_{\sigma}$ over the $10^6$ yr interval, we can adopt it as the initial parameter in the subsequent analysis.

\subsection{The long-term run}

In this subsection we make a complete investigation of the stability of Twotinos on inclined orbits for the age of the Solar system ($4\times10^9$ yr). Among the selected librating particles from the pre-run, we refer to the survivals in the long-term run as ``stable'', and the others that leave the 1:2 NMMR as ``unstable''. Our results are presented on the initial $(A_{\sigma}, i)$ plane at different initial $e$. In the following we will discuss in detail the dynamics regarding the symmetric and asymmetric resonances, respectively.

\subsubsection{The symmetric resonance}
\label{stabilityS}

The stability of the symmetric 1:2 NMMR is explored in the orbital element space ($e=0.05$, $i=0^{\circ}$--$10^{\circ}$), ($e=0.15$, $i=0^{\circ}$--$30^{\circ}$) and ($e=0.25$--0.4, $ i=0^{\circ}$--$90^{\circ}$). For initial horseshoe orbits began with $\sigma=180^{\circ}$, Fig. \ref{sym} summarizes the resonant states for surviving particles after $4\times10^9$ yr of integration time. Squares represent particles in the persistent symmetric resonance (i.e., S), stars denote the alternation between symmetric and asymmetric librations (i.e., S\&A), and crosses denote the transition between libration and circulation (hereafter L\&C). Note that horseshoe orbits are practically stable for inclinations up to  $20^{\circ}$. This upper limit of $i$ for symmetric Twotinos is comparable to the critical value of $15^{\circ}$ given in Lykawka \& Mukai  (2007) and Tiscareno \& Malhotra (2009). 

\begin{figure}
  \centering
  \begin{minipage}[c]{0.6\textwidth}
  \centering
  \hspace{-3cm}
  \includegraphics[width=9cm]{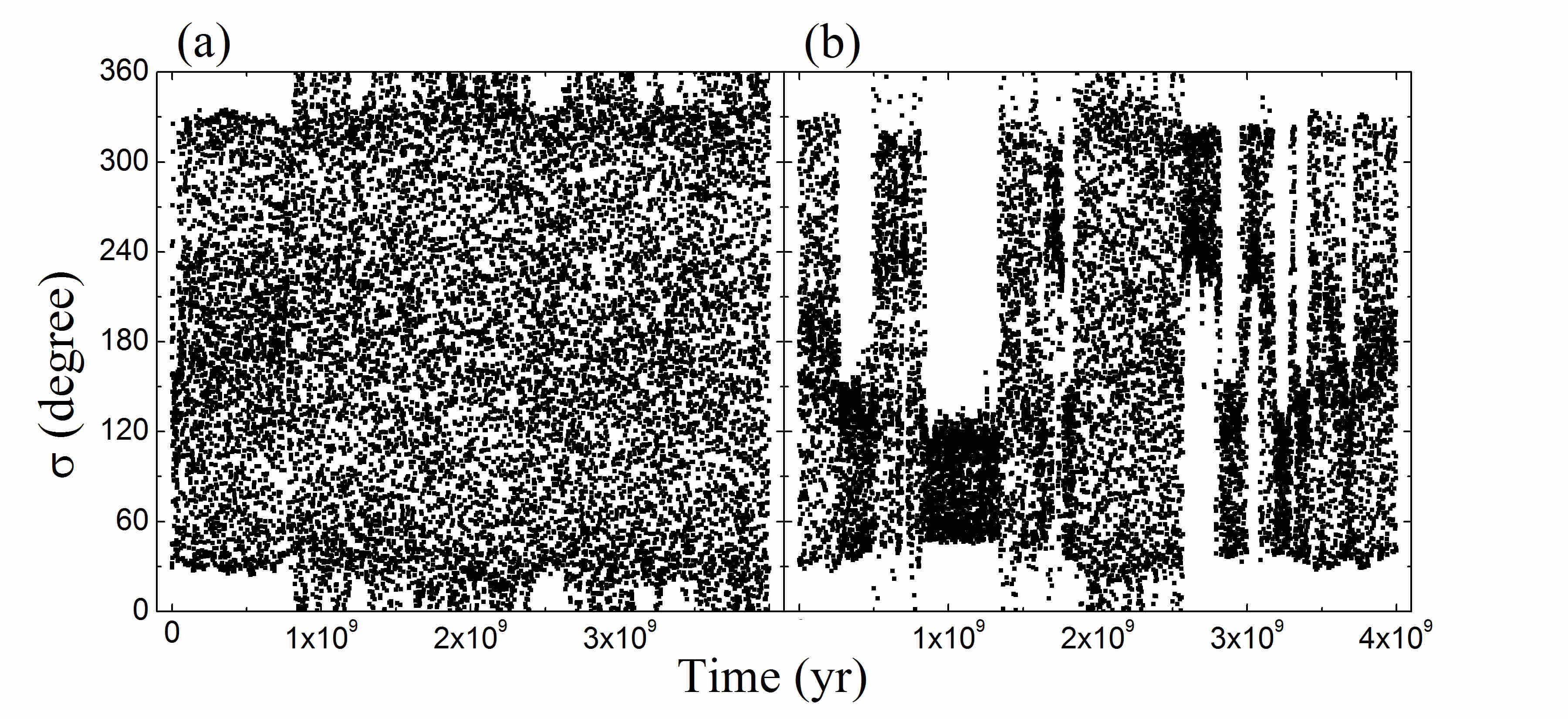}
    \vspace{-0.05cm}
  \end{minipage}
  \begin{minipage}[c]{0.6\textwidth}
  \centering
  \hspace{-3cm}
  \includegraphics[width=9cm]{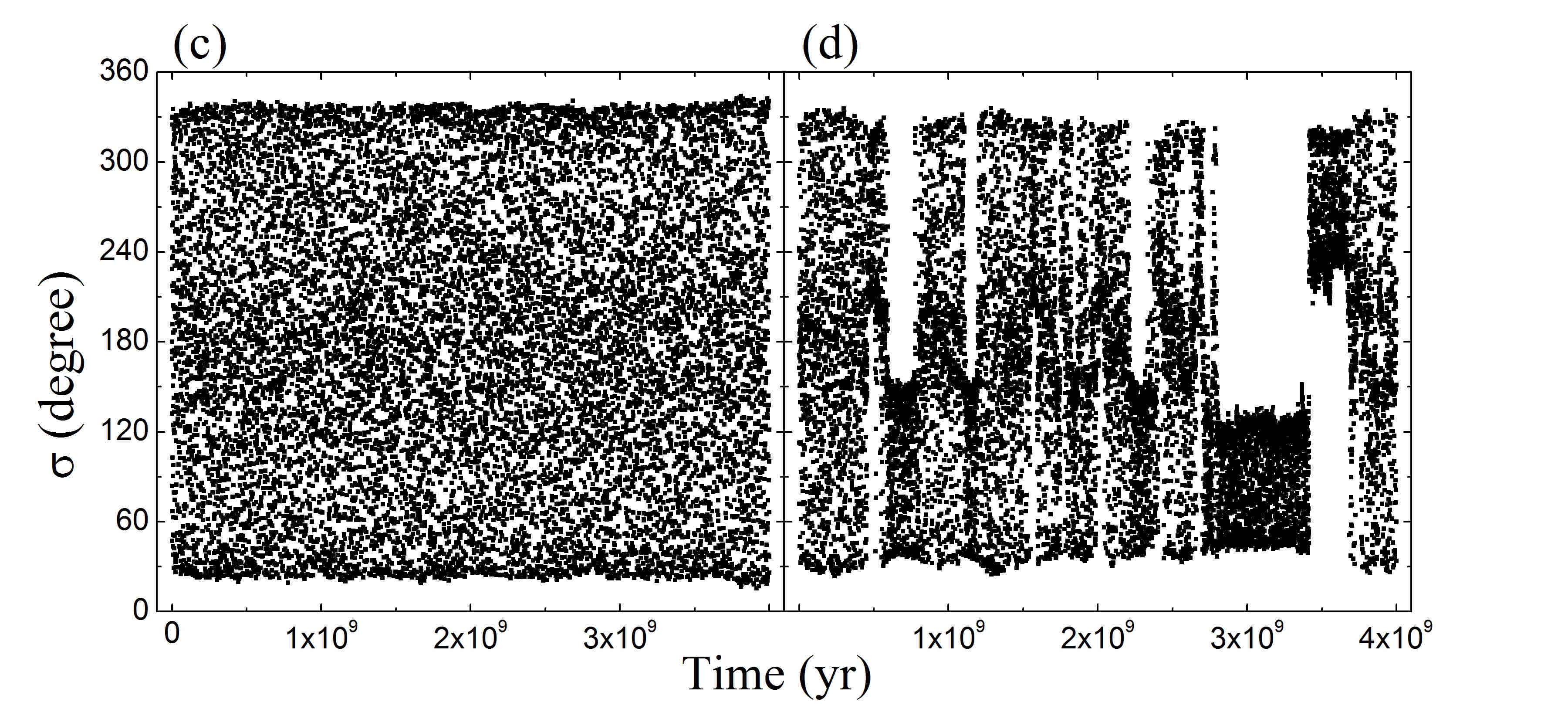}
  \end{minipage}
  \caption{Three typical scenarios for the time evolution of the resonant angle $\sigma$ for the stable Twotinos began on horseshoe orbits: the transition between libration and circulation for initial $i\ge10^{\circ}$ (panels a, b), the persistent symmetric libration (panel c) and the alternation between symmetric and asymmetric librations (panel d) for initial $i=0^{\circ}$.}
 \label{LC}
\end{figure}  

For survivals with initial $i\ge10^{\circ}$, their resonant amplitudes can touch $180^{\circ}$ during the evolution, but such circulation phase is not permanent. We find that these candidate Twotinos are literally on the L\&C trajectories (Fig. \ref{sym}), i.e., close to the separatrix of the 1:2 NMMR. Although the near-separatrix motion is chaotic, some objects may have orbits with diffusion speed too slow to escape from this resonance over the age of the Solar system. Two typical L\&C samples are given in Figs. \ref{LC}a and b, both from the case of initial $i=10^{\circ}$ and $e=0.25$. As a matter of fact, under a long enough integration time on the order of 1 Gyr, all the observed symmetric Twotinos with $i>5^{\circ}$ (see Fig. \ref{observed}) will experience the transition from libration to circulation and vice-versa several times. 

For the case of initial $i=0^{\circ}$, beside the L\&C trajectories, a number of test particles have resonant amplitudes smaller than $180^{\circ}$ throughout the entire simulation. Most of these particles exhibit stable symmetric libration about $\sigma=180^{\circ}$, and another small group follows the frequent alternation between oscillations of  $\sigma$ around the symmetric and asymmetric centres (Fig. \ref{sym}). It can clearly be seen that such S and S\&A particles occupy a wide region for initial $e=0.15$ and 0.25, where the coverage of initial $A_{\sigma}$ exceeds $20^{\circ}$. While for smaller $e=0.05$, the initial horseshoe orbits are more likely to undergo temporary circulation, and there is only one S particle produced in our simulations. This may imply that low-$i$ and high-$e$ orbits are favorable for maintaining the symmetric libration. Examples of S and S\&A trajectories with initial $i=0^{\circ}$ and $e=0.25$ are displayed in Figs. \ref{LC}c and d, respectively.

It must be noted that the resonance-locked configurations for initial symmetric librators is possible only at small $i$. Although we did not go into details of this phenomenon, one conjecture is that the resonance separatrix (outer bolded line in Fig. \ref{phase}) might dissolve into a thick chaotic layer for large $i$ (Malhotra 1996). Consequently, some highly inclined Twotinos could stick to the vicinity of this fuzzy boundary and have their $\sigma$ switching between libration and circulation from time to time. From additional simulations we estimate the maximum $i$ of S and S\&A trajectories to be  $\sim5^{\circ}$. Again, we test this tentative limit on the observed Twotinos in the extended integration up to 1 Gyr and identify an S trajectory with $i=3^{\circ}$ and $e=0.225$.

Overall, the S, S\&A, and L\&C orbits with $i\le20^{\circ}$ and $e\le0.25$ are expected to have been preserved to the present day. In Fig. \ref{sym}, we also see that all the survivals in our simulations have initial $A_{\sigma}<175^{\circ}$, beyond which symmetric librators will lose their stability. Thus this $A_{\sigma}$-criterion can roughly be used to judge the long-term stability of horseshoe orbits in the 1:2 NMMR. 

\subsubsection{The asymmetric resonance}

\begin{figure*}
  \centering
  \begin{minipage}[c]{1\textwidth}
  \hspace{-0.7cm}
  \centering
  \includegraphics[width=15cm]{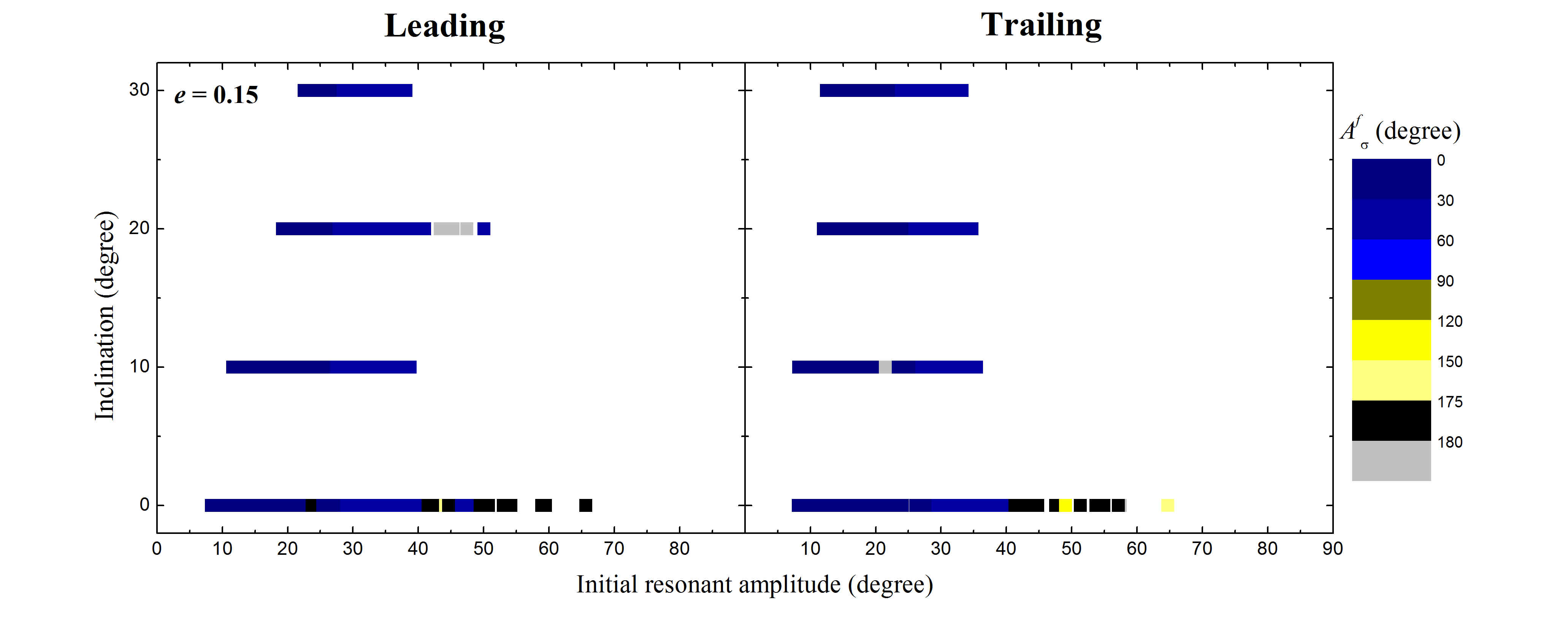}
  \end{minipage}
  \begin{minipage}[c]{1\textwidth}
  \vspace{-0.1 cm}
   \hspace{-0.7cm}
  \centering
  \includegraphics[width=15cm]{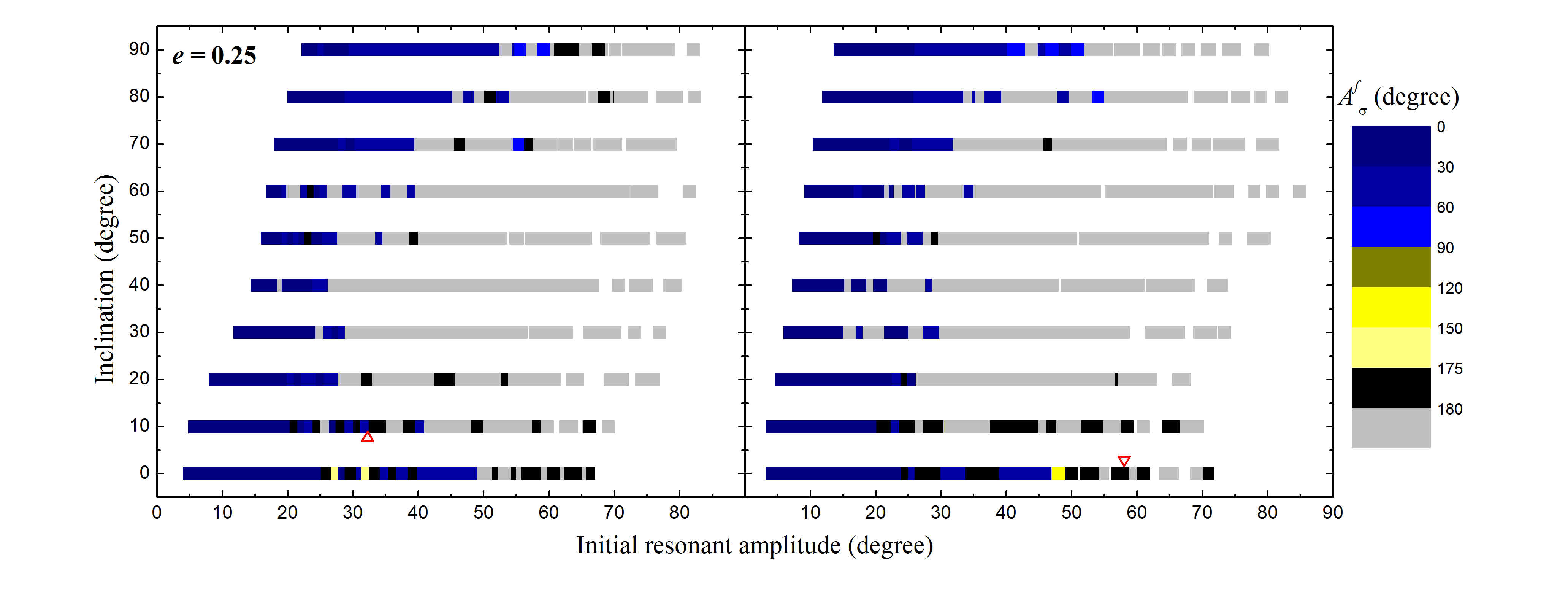}
  \end{minipage}
    \begin{minipage}[c]{1\textwidth}
  \vspace{-0.1 cm}
   \hspace{-0.7cm}
  \centering
  \includegraphics[width=15cm]{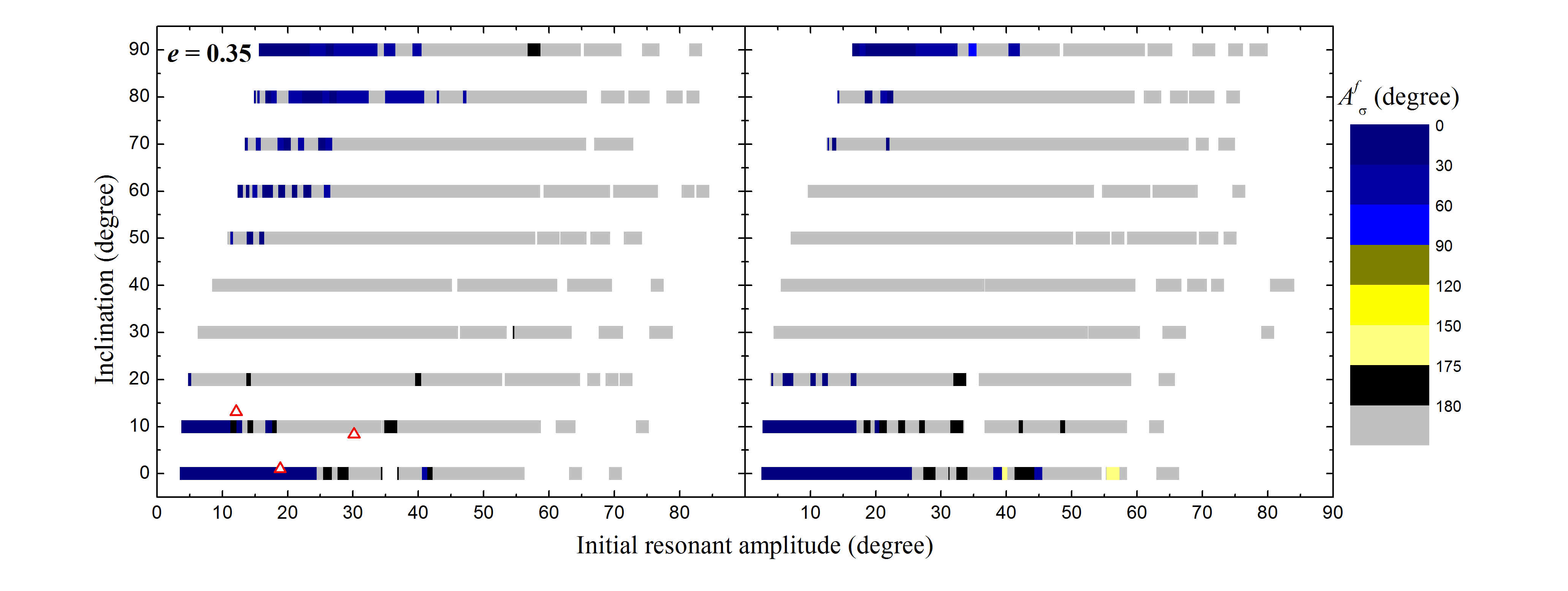}
  \end{minipage}
    \begin{minipage}[c]{1\textwidth}
  \vspace{-0.1 cm}
   \hspace{-0.7cm}
  \centering
  \includegraphics[width=15cm]{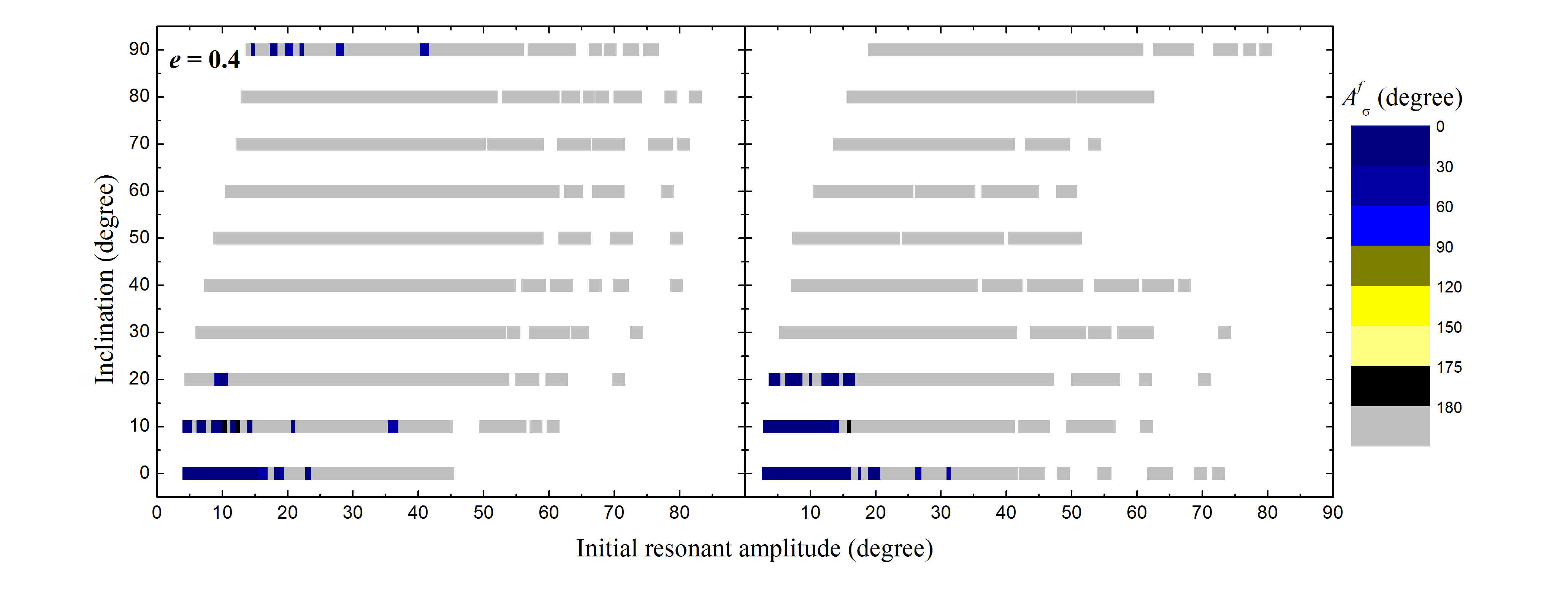}
  \end{minipage}
  \caption{The full resonant amplitudes $A^f_{\sigma}$ measured over the entire 4 Gyr integration for the initial leading (left column) and trailing (right column) librators in the 1:2 NMMR, for initial $e=0.15$, 0.25, 0.35 and 0.4 (top to bottom). The blue strips represent the stable tadpole orbits that can persist for the lifespan of the Solar system. The triangles in the middle two rows indicate the observed asymmetric Twotinos with $e=0.25\pm0.015$ and $e=0.35\pm0.015$, respectively.}
 \label{asym}
\end{figure*}

We now concentrate on the stability of particles in the asymmetric 1:2 NMMR. The initial $(e, i)$ space is the same as that in the symmetric case, but tadpole orbits with $e=0.05$ are discarded as they will turn to the studied S\&A orbits in less than 1 Myr. We show in Fig. \ref{asym} the dynamical maps for the initial leading (left column) and trailing (right column) Twotinos. The coloured strip in these panels indicates for an orbit the full resonant amplitude, $A^f_{\sigma}$, which has been measured over the entire 4 Gyr integration. In the blue region, particles can be stabilized on tadpole orbits for the lifespan of the Solar system. The yellow strips correspond to initial asymmetric librators that went into the horseshoe regime, and black ones correspond to survivals that experienced the temporary circulation. The grey strips represent most unstable asymmetric librators, which have escaped from the Kuiper belt. From the 24 observed Twotinos recorded in Fig. \ref{observed}, five asymmetric objects with $e=0.25\pm0.015$ and $e=0.35\pm0.015$ are selected and superimposed on the middle two rows of Fig. \ref{asym}. We can see that there is just one leading sample lying out of the stable regions, and this inconsistency has been discussed in LZS14.

In Fig. \ref{asym}, we can immediately see that the long-lasting tadpole motion in the 1:2 NMMR may spread over the whole inclination range of $i\le90^{\circ}$, and there is no statistically significant difference in $i$ distribution between the leading and trailing librators. In the subset of $i\le30^{\circ}$, we have reproduced nearly the same dynamical features as described in Tiscareno \& Malhotra (2009), e.g., stable tadpole orbits can always be found but they become less prominent with increasing inclinations. Here, we have further confirmed these results within a longer timespan of 4 Gyr relative to Tiscareno \& Malhotra's work (1 Gyr), and also updated the stability limit of $i$ to $90^{\circ}$. Analogous to LZS14, the possible upper $i$ cut-off of Twotinos will be further constrained by the process of resonant capture and retention, as we shall have done for the sweeping 1:2 NMMR in the next section. 

Of the initial $(e, i)$ groupings we investigated, test particles satisfy the condition $e>e_c(i)$, i.e., the allowance for the asymmetric libration. Then the initial $A_{\sigma}$ can roughly characterize the regular and chaotic tadpole orbits. For particles with small $A_{\sigma}$, they may be stably librating around either of the two asymmetric centres. Since $A^f_{\sigma}$ was designed to measure the maximum variation of $\sigma$ over the 4 Gyr interval, those particles with $A^f_{\sigma}<90^{\circ}$ (blue region in Fig. \ref{asym}) would never undergo the transformation between the leading and trailing motions, and they exhibit the most regular behavior of asymmetric Twotinos. While for tadpole particles began with $A_{\sigma}>50^{\circ}$, they are generally unstable and leave the asymmetric 1:2 NMMR in our numerical simulation. This instability limit of resonant amplitudes is consistent with that of Tiscareno \& Malhotra (2009) for $i\le35^{\circ}$. Therefore, the criterion of $A_{\sigma}<50^{\circ}$ could serve as useful diagnostics for the long-term stability of all prograde tadpole orbits in the 1:2 NMMR.

For test particles with initial $i=0^{\circ}$, a few of them diffuse out of the asymmetric libration zone and into horseshoe orbits. After that, they could stay on the S or S\&A trajectories and end with $90^{\circ}<A^f_{\sigma}<175^{\circ}$, as shown by yellow strips in Fig. \ref{asym}. Note that such transferring routes do not exist at large $i$ ($\ge10^{\circ}$). From this case, we again prove that the high-$i$ Twotinos would never be observed in the S or S\&A orbits for a long enough time.

One can also find in Fig. \ref{asym} that, the minimum resonant amplitude of Twotinos could approach zero when $i=0$ and tends to larger value with increasing $i$. As we analyzed in LZS14, the low limit of $A_{\sigma}$ is predicted by the variation of the SLC, which becomes more pronounced for higher $i$. This peculiar resonant feature is related to a robust determination of the stability of inclined tadpole orbits. If the dependence of the asymmetric libration centre on the inclination was ignored, test particles would begin with a value of $\sigma$ that displaced away from the real  $\sigma^{L}_0$ ($\sigma^{T}_0$). It means that the minimum initial $A_{\sigma}$ of these particles would be even larger, thereby they might be beyond the stable (blue) bands and reside in the totally unstable (grey) region, especially for highly inclined orbits (e.g., at $i=20$ for $e=0.4$).

For the case of moderate $e$ ($\sim0.15$), the dynamical structure of the asymmetric 1:2 NMMR is fairly simple (Fig. \ref{asym}, top panels). It can be seen that the long-lived tadpole orbits are prevalent at any $i$ ($\le30^{\circ}$), and they almost cover the whole region of initial $A_{\sigma}<40^{\circ}$. From our detailed examination, we found that the orbital elements $e$, $i$ and $A_{\sigma}$ of stable tadpole librators would barely change over 4 Gyr. This suggests a most regular place in eccentricity. Nevertheless, for $e=0.15$, one must recall a vacuum that the asymmetric libration has been precluded at $i>35^{\circ}$.

For the case of large $e$ ($\sim0.25$--0.4), although the asymmetric libration is theoretically allowed for any prograde orbit, the stable regions of asymmetric resonance are remarkably different with the change of $i$ (Fig. \ref{asym}, below top panels). In general, the stability limit of initial $A_{\sigma}$ has a maximum value of $\sim50^{\circ}$, while it decreases first with the increase of $i$ and then raises instead when $i\gtrsim50^{\circ}$. We found that the destabilization of tadpole orbits is principally accounted for by the Kozai mechanism (the libration of the argument of perihelion $\omega$ of a planetesimal):

(1) At $i=0^{\circ}$, obviously an inclination decrease is impossible, so the eccentricities of Kozai Twotinos are forced to decrease towards 0 ($<e_c$). As a result, the two islands of asymmetric librations join together again to form a single island of symmetric librations in the phase space (Beaug\'{e} 1994; Malhotra 1996). Then the transition from tadpole to horseshoe motion is certain to take place. Afterwards, these Twotinos with small $i$ would undergo very similar horseshoe-type evolutions as we showed in Section 3.2.1: some of them are stably librating on S or S\&A trajectories (yellow strips), some could survive on L\&C trajectories to the end of the 4 Gyr integration (black strips), and still others become unstable (grey strips). 

(2) For tadpole orbits with $i=10^{\circ}$--$20^{\circ}$, besides the $e$-decrease trend as in the case of $i=0^{\circ}$, the Kozai mechanism can also bring some objects to more eccentric mode with the cost of $i$ being lower. In such event the symmetric and asymmetric zones can co-exist (Fig. \ref{phase}), passing through their $\infty$-type separatrix, initial tadpole particles may travel into horseshoe orbits. However, in this interval of high inclinations, symmetric librators will finally achieve $A^f_{\sigma}=180^{\circ}$, and thus they can only be stabilized on L\&C trajectories. It is worth mentioning that, among asymmetric librators with the highest final eccentricities ($>0.37$), a handful of them could be stable Kozai Twotinos (also see the observed samples in Fig. \ref{observed}). 

(3) When $30^{\circ}\le i \le90^{\circ}$, the unstable tadpole particles evince the libration of $\omega$ and then leave the 1:2 NMMR without an intermediate stage in the symmetric resonance zone. We notice that few tadpole librators may maintain relatively steady resonant amplitudes less than $<90^{\circ}$ on time-scales of Gyrs, while they go into circulation near the very end of the simulation. Due to the complete absence of L\&C trajectories at $i\ge30^{\circ}$, such circulating survivals (colour black) would be cleared away soon and are of no interest for potential Twotinos.

Apart from the destabilizing effect of the Kozai mechanism, we have not detected any other secular dynamics responsible for the unstable tadpole orbits. Actually, Morbidelli et al. (1995) argued that the $\nu_8$ and $\nu_{18}$ secular resonances (see LZS14 for definitions) cannot overlap the 1:2 NMMR, since the changes of the longitude of perihelion and the longitude of ascending node of a Twotino are too slow to match those of the planets.

%________________________________________________________________________________________________________________________________________________________

\section{Resonant capture and orbital evolution}

As in LZS14, we numerically simulated the evolution of primordially inclined planetesimals under the classical migration scenario of Malhotra (1995), in which Neptune originates from $\sim$23.2 au and moves slowly to its current location at $\sim$30.2 au. The migration time-scale $\tau$ is still chosen to be $2\times10^7$ yr to fulfill the adiabatic invariant condition (Melita \& Brunini 2000). Unlike for the Plutinos, the formation of the Twotinos within the planetary migration model has been studied by very few authors (Chiang \& Jordan 2002; Murray-Clay \& Chiang 2005).

In this paper, we focus on the dependence of the capture efficiency of the 1:2 NMMR on planetesimal's initial inclination $i_0$. Originally, test particles are assumed to be in the region $37.7\le a \le46.8$ au, where 1 au exterior to the initial location of the 1:2 NMMR and 1 au interior to the final 1:2 NMMR. This very distribution of $a$ is inspired by Chiang \& Jordan (2002), for the sake of ideally allowing all particles to be trapped into the sweeping 1:2 NMMR. Then, we designed a series of runs for test particles with $i_0$ ranging from $0^{\circ}$ to $90^{\circ}$ in steps of $10^{\circ}$. For each $i_0$ in a run, the 911 test particles with $\Delta a=0.01$ au are introduced. Their initial eccentricities are chosen to be 0.01, which is smaller than the critical value of 0.06 that would make the 1:2 NMMR fail to capture planetesimals (Melita \& Brunini 2000). The other three orbital elements $\Omega$, $\varpi$ and $\lambda$ are chosen randomly between 0 and $2\pi$.

In the numerical orbital calculations that include the planet migration, we also employ the SWIFT\_RMVS3 integrator with a time-step of 0.5 yr. We integrate the system for $10^8$ yr, that is 5 times the assumed value of $\tau$ and long enough for planets to reach their current configurations. To better describe the results, for particles survived inside the 1:2 NMMR, their resonant amplitudes $A_{\sigma}$ have been measured over the last $10^6$ yr in the integrations.

According to the obtained stability criterions based on $A_{\sigma}$, we are able to roughly determine whether a resonant orbit is stable enough to be included in the final statistics. Thus we will refer to the potentially long-lived Twotinos as: captured symmetric Twotinos (CSTs) with $A_{\sigma}<175^{\circ}$; captured leading Twotinos (CLTs) and captured trailing Twotinos (CTTs) with $A_{\sigma}<50^{\circ}$. These three populations are collectively named ``captured Twotinos''. It is plausible that some fraction of captured Twotinos on highly inclined orbits may walk out of the 1:2 NMMR very slowly, as we discuss later.

\subsection{Capture into the 1:2 NMMR}

\begin{table}
\hspace{-0.5cm}
\centering
\begin{minipage}{8.3cm}
\caption{The statistics of CSTs for initial inclinations $i_0$ from $0^{\circ}$ to $90^{\circ}$ in the $10^8$ yr migration simulation. For each $i_0$, 911 test particles with initial $37.7\le a \le 46.8$ au and $e=0.01$ were used to derive the capture efficiency $R_S$, the number of Kozai Twotinos $K_S$, the ranges of final $e$ and $i$, and the minimum resonant amplitude Min$A_{\sigma}$. For Kozai CSTs, the number in parentheses refers to the event that the libration of $\omega$ can persist for another $10^8$ yr (see Section 4.2 for details).}      % title of Table
\label{CST}
\begin{tabular}{c c c c c c}        % centered columns (9 columns)
\hline                 % inserts double horizontal lines
$i_0(^{\circ})$     &         $R_S$ (per cent)      &        $K_S$             &               $e$                  &            $i(^{\circ})$               &        Min$A_{\sigma}(^{\circ})$     \\
 
\hline

              0               &                    4.7                     &        1(1)               &           0.04--0.34            &                    0--16                   &                        124                                \\
    
            10              &                     4.3                     &          0                  &           0.05--0.34            &                    5--19                   &                        119                                 \\
        
            20              &                     0.5                     &          0                  &           0.06--0.16            &                   20--23                  &                        138                                   \\
            
            30              &                     0.2                     &          0                  &           0.32--0.34            &                   25--28                   &                       159                                    \\
   
         40--90          &                       0                       &          --                  &                  --                     &                      --                         &                           --                                       \\

\hline
\end{tabular}
\end{minipage}
\end{table}

For the capture into the symmetric 1:2 NMMR, the CSTs with $i_0$ in the range of $0^{\circ}-30^{\circ}$ are always present, as shown in Table \ref{CST}. Out of 911 test particles at each $i_0$, the capture efficiency $R_S$ could exceed 4 per cent for $i_0\le10^{\circ}$, while it decreases to less than $\sim$0.5 per cent for $i_0=20^{\circ}, 30^{\circ}$. We interpreted such sharp decrease of $R_S$ as a consequence of the weak retainment of high-$i$ symmetric librators. As the results illustrated in Fig. \ref{sym}, the concentration of Twotinos is significant at $i\le10^{\circ}$ and becomes quite sparse at higher $i$ ($\sim20^{\circ}$). It should also be noted that stable horseshoe motion does not exist for $i\ge30^{\circ}$, this instability limit may provide strong quantitative constraint on the final $i$ of CSTs. In fairly good agreement, Table \ref{CST} shows that all the CSTs have $i\le28^{\circ}$, and there is a complete absence of more inclined bodies having $i_0\ge40^{\circ}$. 

Among the 43 ($911\times4.7\%$) CSTs from the case of $i_0=0^{\circ}$, 38 remain on low inclination orbits ($i<5^{\circ}$) when the migration has stopped.  These objects populate the most regular zone of symmetric librations, and they may settle on S or S\&A trajectories with $A_{\sigma}$ never reaching $180^{\circ}$ over the age of the Solar system. While for the 46 ($911\times(4.3+0.5+0.2)\%$) CSTs with $i_0=10^{\circ}-30^{\circ}$, their inclinations can hardly attain values below $5^{\circ}$. Thus they will eventually experience the (temporary) circulation phase and can only be stabilized on L\&C trajectories. From this information, we estimate a likelihood probability of $38/(43+46)\approx43\%$ for symmetric Twotinos to stay on persistently librating, resonance-locked orbits.

\begin{table*}
\centering
\begin{minipage}{13cm}
\caption{The same as Table \ref{CST}, but for CLTs and CTTs with subscripts ``L'' and ``T'', respectively.}      % title of Table
\label{CLTandCTT}
\begin{tabular}{c c c c c c c c}        % centered columns (9 columns)
\hline                 % inserts double horizontal lines
$i_0(^{\circ})$   &      $R_L$ (per cent)     &    $R_T$ (per cent)    &      $K_L$      &        $K_T$        &         $e$                 &      $i(^{\circ})$    &     Min$A_{\sigma}(^{\circ})$     \\

\hline

              0             &                  26.2               &                   23.2             &        3(2)       &          2(0)         &      0.08--0.34         &            0--12           &                    6                                \\
    
            10             &                  21.6               &                   19.5             &        1(1)        &            0           &      0.07--0.37         &            1--19           &                    5                                 \\
        
            20             &                   6.6                &                   11.0             &         2(0)        &           0           &       0.08--0.40        &             9--24          &                   11                                 \\
            
            30             &                   2.0                &                    2.0              &         1(0)        &            0           &       0.09--0.39        &            21--32         &                  11                                   \\
            
\hline \hline
   
           40             &                    0.7                &                    0.2              &           0           &           0            &      0.06--0.32         &            38--41         &                    8                                   \\
           
           50             &                    0.2                &                    0.2              &           0           &           0            &       0.26--0.37        &            47--49         &                  17                                  \\
           
           60             &                      0                 &                      0                &           --           &          --            &              --                 &                 --             &                   --                                   \\
           
           70             &  (1/911)$\times100$   &                      0               &           0           &           --           &            0.26              &               68             &                  25                                   \\
           
           80             &                     0                  & (1/911)$\times100$  &           --           &           0           &            0.27              &               78             &                  42                                   \\
             
           90             &   (1/911)$\times100$ &                      0                &           0           &           --           &           0.32               &               88             &                  38                                   \\         
   
\hline
\end{tabular}
\end{minipage}
\end{table*}

A comprehensive information of CLTs and CTTs is given in Table \ref{CLTandCTT}. For test particles with $i_0=0^{\circ}$, the capture efficiency of the asymmetric 1:2 NMMR (i.e., the sum of both leading and trailing cases) is $R_L + R_T\approx49.4$ per cent, which is a bit higher than that in Chiang \& Jordan (2002) ($\sim44.5$ per cent). Considering their adopted migration time-scale of $10^7$ yr, such an efficiency enhancement is obviously due to a slower migration here with $\tau=2\times10^7$ yr. More importantly, we find that there are more CLTs (26.2 per cent) than CTTs (23.2 per cent), contrary to the difference in the numbers of these two populations in Chiang \& Jordan's calculations. 

Assuming a planar model, the SLC of 1:2 resonant orbits is fixed at $180^{\circ}$ for $e\sim0.01$ where $e_a(=0)<e<e_c(=0.037)$. This means that only the symmetric resonance trapping is possible at the moment of 1:2 NMMR encounter. As time passes, a particle may transit from symmetric to asymmetric libration when its eccentricity has been pumped up via the sweeping 1:2 NMMR or the associated Kozai mechanism. As pointed out before by Murray-Clay \& Chiang (2005), for $\tau\le10^7$ yr, symmetric librators are preferentially caught into the tailing rather than the leading asymmetric resonance during the outward migration; however, this trend could be reversed in a migration model with larger $\tau$. The spatial distribution for which more asymmetric Twotinos lie in the leading island is also supported by Lykawka \& Mukai (2007).

With the increase of $i_0$, the probability of capture into the asymmetric 1:2 NMMR would be modified dramatically. Firstly, for test particles with $i_0\le50^{\circ}$, the capture efficiency $R_L + R_T$ fails from 49.4 per cent for $i_0=0^{\circ}$ to 0.4 per cent for $i_0=50^{\circ}$. Particles that originate in this $i_0$-range can reside deeply in the asymmetric islands (with low Min$A_{\sigma})$, and therefore become long-term residents of the 1:2 NMMR. Here we would like to stress that an overwhelming majority of CLTs and CTTs have final $e$ larger than $e_c$ with regard to their particular $i$. But there are two exceptions coming from the case of $i_0=40^{\circ}$, because they have considerable oscillations of $e$ and will not strictly follow our semi-analytical approximation. Secondly, for each $i_0\ge60^{\circ}$, at most only one of the 911 test particles can be hijacked into the asymmetric 1:2 NMMR. For the two CLTs from $i_0=70^{\circ}$, $90^{\circ}$ and the one CTT from $i_0=80^{\circ}$ with relatively large $A_{\sigma}$ (= Min$A_{\sigma}$), they are actually embedded in or very close to the unstable (grey) area in Fig. \ref{asym}. By extending the integrations, we confirm that these three individuals will evolve into the L\&C orbits in less than several hundreds of Myrs, and then escape from the resonance. 

On the other hand, the influence of high $i_0$ could be very strong on the ratio of CLTs to CTTs. An inclined particle captured by the sweeping 1:2 NMMR first evince the oscillation of $\sigma$ around the SLC, which may locate at any position between $0^{\circ}$ and $360^{\circ}$ when $e<e_a(i)$. As for the case of $i_0=10^{\circ}$, on extremely short time-scales, the value of $e$ for the captive would be excited above $e_a(i=10^{\circ})=0.038$ by this resonance, then it will fall into the typical symmetric libration around the GLC at $180^{\circ}$. The subsequent evolution of such symmetric librators is equivalent to the scenario in the planar model described above, producing more Twotinos in the leading island. However, when $i_0\ge20^{\circ}$, the captive would never undergo a period of symmetric libration around $\sigma=180^{\circ}$ prior to the emergence of the asymmetric islands. This is because, for orbits with $i>15^{\circ}$, the GLC at $180^{\circ}$ does not exist when $e<e_c(i)$, as we argued in Section 2. As a result, the probability of the captive to enter the leading or trailing island cannot be determined by the analytic methods used in Murray-Clay \& Chiang (2005). Our numerical experiments demonstrate that the relative population of the CLTs and the CTTs is continuously varying with $20^{\circ} \le i_0 \le 50^{\circ}$.

\begin{figure}
 \hspace{-0.5cm}
  \centering
  \includegraphics[width=9cm]{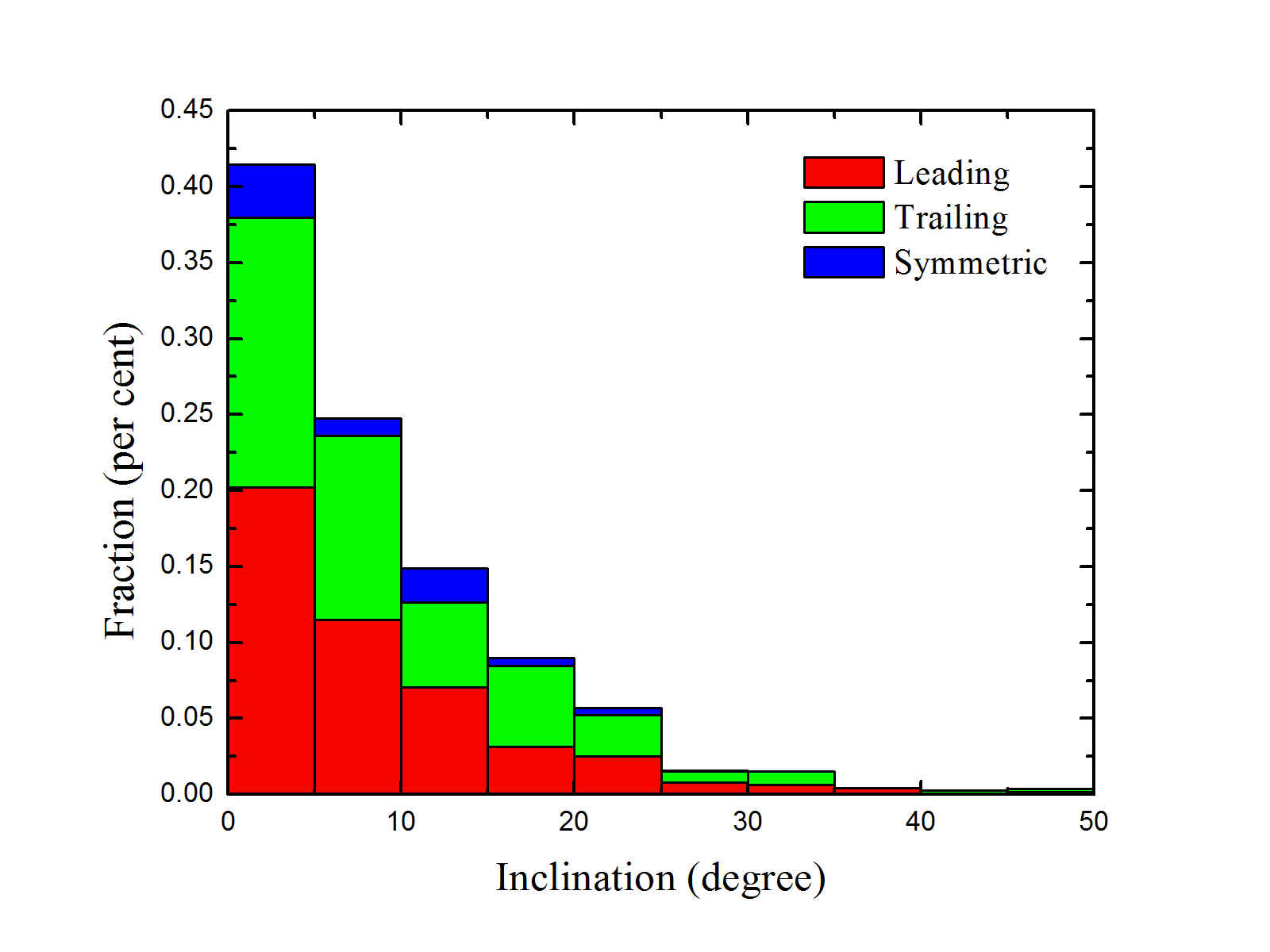}
  \caption{Cumulative histogram of the inclination distribution of captured Twotinos from the migration simulation. The red, green and blue histograms show the fraction of the leading, trailing and symmetric librators, respectively.}
  \label{distribution}
\end{figure}

In view of the above, the mechanism of resonant capture may exclude the existence of Twotinos with $i>50^{\circ}$. Fig. \ref{distribution} supplies the distribution of inclinations for CLTs (red), CTTs (green) and CSTs (blue), at the end of the $10^8$ yr migration simulation. It is visually apparent in this figure that most objects have inclinations within the range of the observed values ($i<25^{\circ}$). For even inclined Twotinos, the relative fraction is estimated to be about 4 per cent. This number is considered to be deduced from the total inclination distribution of all potential Twotinos in the Kuiper belt. As most surveys for KBOs target near the ecliptic plane, the probability of discovering an object is crudely proportional to $1/\sin{i}$ (Brown 2001). Thus, migration models can predict a weighted fraction as low as $\sim$0.1 per cent for the unknown Twotinos with $i>25^{\circ}$ in future discoveries.

Nevertheless, a more accurate comparison between the inclination distribution of observed Twotinos (Fig. \ref{observed}) and that produced in our simulations (Fig. \ref{distribution}) shows a noticeable discrepancy. If we consider the high-inclination population with $i>10^{\circ}$, the real Twotinos give a fraction of $\sim40\%$, while the captured ones weighted with $1/\sin{i}$ yield only $\sim2\%$. Moreover, under a theoretical point of view, we chose the extreme and unrealistic disk initially having a uniform distribution in inclination up to $90^{\circ}$. This may be an artifact in a primordial planetesimal disk, where would likely to have existed much less particles with higher inclinations. We expect that the intrinsic percentage of captured Twotinos with $i>10^{\circ}$ should be even smaller, and too low to account for current observations. Consequently, the Twotinos must not have been formed solely by the resonance sweeping mechanism. An alternative hypothesis is that, a portion of Twotinos may have originated from the chaotic capture of scattered KBOs, resulting in a larger fraction of inclined Twotinos (Lykawka \& Mukai 2006). 

The last piece of information about the migrating evolution of captured Twotinos is linked to the eccentricity distribution. We notice in Tables \ref{CST} and \ref {CLTandCTT} that a few particles are transported to nearly circular orbits with $e$ as small as 0.04. In the context of adiabatic model, the longer migration distance leads to larger eccentricity excitation of a planetesimal locked in the 1:2 NMMR. Given the outer edge of the planetesimal disk at 46.8 au, from equation (9) in LZS14, it follows that a Twotino with $i<25^{\circ}$ should have $e\gtrsim0.1$. The lower eccentricities obtained by our simulations can be readily accounted for by the temporary Kozai mechanism, which may last shorter than a complete cycle of $\omega$-libration and transfer $e$ to $i$ irreversibly during the planet migration era (Gomes 2003).

\subsection{Kozai mechanism}

Of captured Twotinos coming from our simulations, many objects exhibit the libration of $\omega$ after entering the migrating 1:2 NMMR, but only a small subset occupies the Kozai mechanism at the end of the integration, recorded in Tables \ref{CST} and \ref {CLTandCTT}. Were the final 10 $\omega$-librators lasting one to several Kozai cycle periods also transient residents? To better confirm their Kozai behaviors, we continued the integrations of these orbits for another $10^8$ yr without any migration. There are a total of 4 remaining $\omega$-librators: 1 in the symmetric resonance, and 3 in the leading asymmetric resonance. 

\begin{figure}
 \hspace{0cm}
  \centering
  \includegraphics[width=8.5cm]{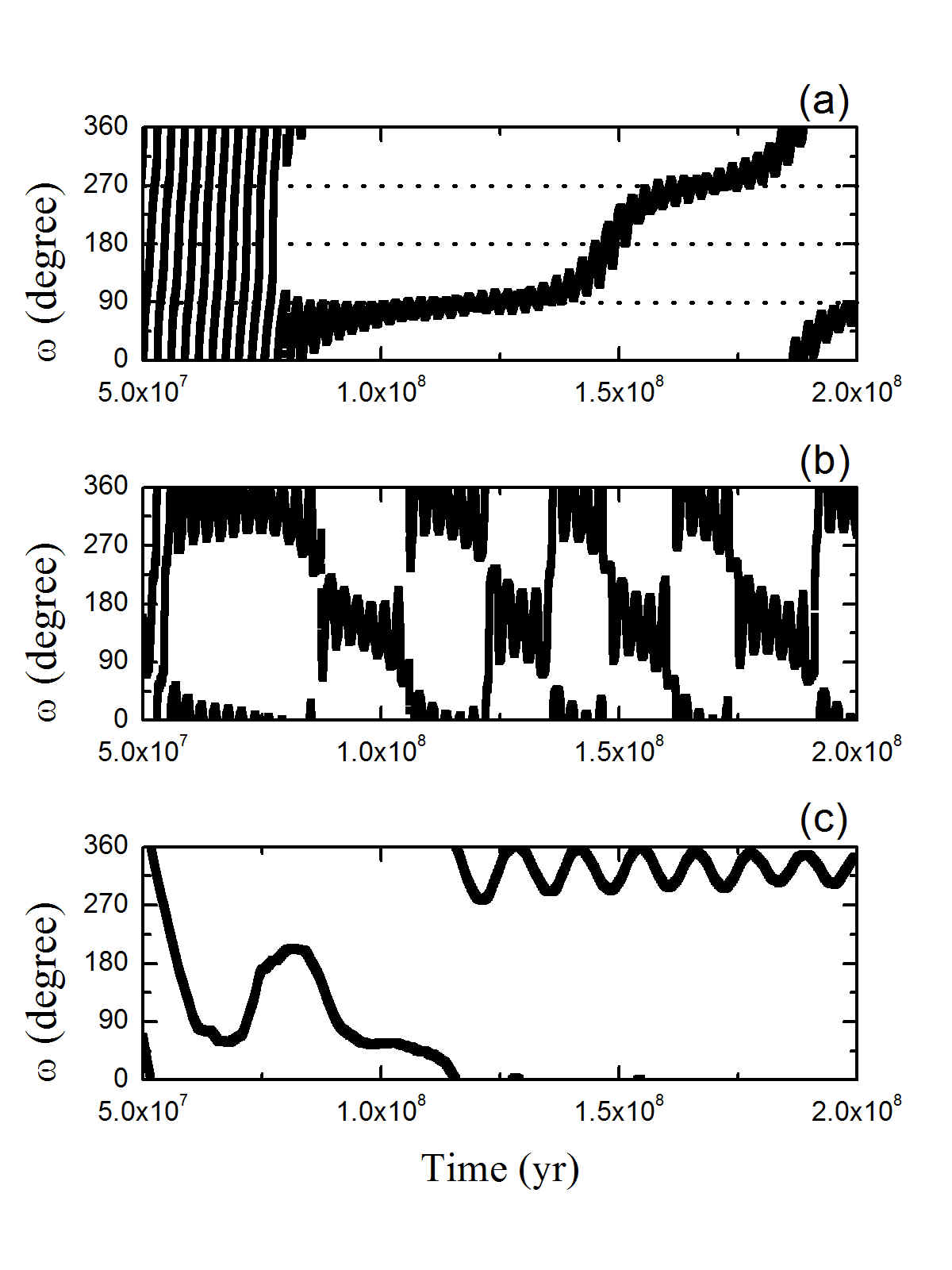}
  \caption{The time evolution of $\omega$-librators in the symmetric (panel a) and the leading asymmetric (panels b, c) 1:2 NMMR.}
  \label{kozai}
\end{figure}

Fig. \ref{kozai}a displays the time evolution of $\omega$ for the one and only CST associated with the stable Kozai state, from the case of $i_0=0^{\circ}$. As it can be seen, the libration centre $\omega_0$ alters from time to time about $0^{\circ}$, $90^{\circ}$, $180^{\circ}$ and $270^{\circ}$, which are equilibrium points of the Kozai mechanism inside the symmetric 1:2 NMMR as shown in Wan \& Huang (2007). These authors also found that such long-lived $\omega$-librator does exist near the separatrix of Kozai islands, and it would spend more time around the stable equilibrium points $\omega_0=90^{\circ}$ and $270^{\circ}$ than around the unstable ones $\omega_0=0^{\circ}$ and $180^{\circ}$ (personal communication).

For the asymmetric 1:2 NMMR, the Kozai dynamics could be much different. Because the libration centre of $\sigma$ is not at $0^{\circ}$ or $180^{\circ}$, the secular and 1:2 resonant terms containing $\sin{\sigma}$ in the disturbing function cannot be averaged to be 0. Gallardo et al. (2012) shows that the Kozai centres $\omega_0$ would be shifted from the above four values. Figs. \ref{kozai}b and c samples two Kozai CLTs from the cases of $i_0=0^{\circ}$ and $10^{\circ}$, respectively. Their $\omega$ may librate around the centres located at $140^{\circ}$ or $320^{\circ}$, which is a difference of $50^{\circ}$ from the stable $\omega_0$ for the symmetric Twotinos. However, the explicit correlation between the $(e, i)$ pair and $\omega_0$ for the asymmetric Twotinos cannot be concluded here, and a more detailed exploration is reserved.

We further note that for the 3 CLTs with $i_0\ge20^{\circ}$ and the 2 CTTs experiencing the Kozai dynamics at $10^8$ yr, none of them can keep on the libration of $\omega$ in our extended run up to $2\times10^8$ yr. This outcome might explain today's observed void of Kozai Twotinos having $i>15^{\circ}$ or residing in the trailing resonance (Fig. \ref{observed}). 

However, a principle issue is that, of the 1075 captured Twotinos with $i<25^{\circ}$, there are only 4 possible $\omega$-librators. This relative fraction of $\sim0.4$ per cent is much too low by contrast to that of currently observed samples ($2/24\approx8.3$ per cent). As argued in the end of Section 4.1, the migration-capture scenario would not be the sole mechanism to produce Twotinos. Here, the deficit of captured Twotinos experiencing the Kozai mechanism gives an extra evidence of this fact.

%______________________________________________________________

\section{Conclusions and discussion}

To continue our previous study of the high-inclination KBOs, in which we explored the dynamic of Plutinos with $i$ up to $90^{\circ}$ (in LZS14), this second paper is devoted to Twotinos. We are acutely aware of the existence of symmetric and asymmetric motions in the 1:2 NMMR, thus Twotinos may furnish a unique window on the distant past of the outer Solar system.

Using the semi-analytical method developed in LZS14, for the first time, we provide a global picture of the dynamical aspects of the 1:2 NMMR in three dimensions. Depending on the critical eccentricities $e_a(i)$ and $e_c(i)$ as functions of the inclination $i$, three different behaviors of the resonant angle $\sigma$ have been found: (1)  if $e<e_a$, the circulation of $\sigma$ by $360^{\circ}$ is certain; (2) if $e_a<e<e_c$, there appear a libration of $\sigma$ around $180^{\circ}$ for $i<15^{\circ}$, and another libration around $0^{\circ}$ for larger $i$; and (3) if $e>e_c$, the asymmetric libration of $\sigma$ becomes possible, meanwhile the symmetric libration around $180^{\circ}$ is also allowed. At zero inclination, we have $e_a=0$ and $e_c=0.037$, equally consistent with earlier works (Beaug\'{e} 1994; Malhotra 1996). But with increasing $i$, these two quantities both shift to larger values, and $e_c$ achieves a maximum of $\sim0.21$ when $i>60^{\circ}$. These characteristics indicate that, for highly inclined orbits, the asymmetric libration islands can only emerge at large eccentricities. Accordingly, we calculated the libration centres $\sigma_0$ for 1:2 resonant orbits with $0^{\circ}\le i \le 90^{\circ}$, at specific $e$. We have shown that, for high-inclination Twotinos, their $\sigma_0$ are strongly dependent on $i$ and could be very different from those of low-inclination ones.

With the initial $\sigma=\sigma_0$ on a particular grid of $(e, i)$ based on the above semi-analytical predictions, test particles initialized on 1:2 resonant orbits were numerically integrated with four Jovian planets for times up to the age of the Solar system. Our principal findings for the long-term stability of Twotinos can be briefly summarized as follows. 

(1) In the symmetric resonance, we find that the stable horseshoe orbits have $e\le0.25$, $i\le20^{\circ}$ and resonant amplitudes $A_{\sigma}<175^{\circ}$. The most regular motion only exists in the low inclination region of $i\lesssim5^{\circ}$, where initial horseshoe orbits can exhibit persistent libration of $\sigma$ throughout the entire 4 Gyr simulation. For more inclined particles, they are fated to undergo the frequent alternation between libration and circulation during the evolution, while many of them can survive as symmetric librators in the 1:2 NMMR.

(2) In the asymmetric resonance, we find that the inclination coverage of stable tadpole orbits is $\le30^{\circ}$ for moderate $e=0.15$, and as high as $90^{\circ}$ for large $e$ between 0.25 and 0.4. By measuring the full resonant amplitude $A^f_{\sigma}$ over the 4 Gyr interval, the dynamical maps have been constructed on the plane of initial resonant amplitude versus initial inclination. It is shown that the stable regions are generally delimited by initial $A_{\sigma}<50^{\circ}$, but their widths are quite distinct at different initial $i$. The unstable tadpole librators are principally accounted for by the appearance of the Kozai mechanism. However, their escaping routes from the 1:2 NMMR could be drastically different for small ($\sim0^{\circ}$), medium ($10^{\circ}$--$20^{\circ}$) and high ($30^{\circ}$--$90^{\circ}$) inclinations. Overall, there is no statistically significant discrepancy in the dynamical structure between the leading and trailing asymmetric resonances.

Then, we further investigated the evolution of primordially inclined planetesimals with respect to the 1:2 NMMR sweeping and capture, for which Neptune starts at $\sim23.2$ au and migrates slowly with a time-scale of $\tau=2\times10^7$ yr. At each $i_0$ between $0^{\circ}$ and $90^{\circ}$ ($\Delta i_0=10^{\circ}$), 911 test particles with $37.7\le a \le 46.8$ au ($\Delta a=0.01$ au) and $e=0.01$ have been introduced. Of surviving objects in the 1:2 NMMR zone at the end of the migration, according to the above stability analysis, we defined captured Twotinos as:  captured symmetric Twotinos (CSTs) with $A_{\sigma}<175^{\circ}$; captured leading Twotinos (CLTs) and captured trailing Twotinos (CTTs) with $A_{\sigma}<50^{\circ}$. 

Our results show that the capture efficiency of the symmetric 1:2 NMMR, $R_S$, decays from 4.7\% to 0.2\% as the initial inclination $i_0$ increases from $0^{\circ}$ to $30^{\circ}$. While for $i_0\ge40^{\circ}$, there is a complete absence of CSTs. The final inclinations of CSTs suggest an upper limit of $i\sim28^{\circ}$ for potential symmetric Twotinos, which is just inside the border of the totally unstable horseshoe region at $i\ge30^{\circ}$. 

As for the asymmetric 1:2 NMMR, the capture efficiency $R_L+R_T$ fails from 49.4 per cent for $i_0=0^{\circ}$ to 0.4 per cent for $i_0=50^{\circ}$. Particles originated in this $i_0$-range can possess small enough $A_{\sigma}$ to reside deeply in the asymmetric islands, and therefore form long-lived residents. For few CLTs and CTTs with relatively large $A_{\sigma}$ from the cases of $i_0\ge60^{\circ}$, they are actually embedded in or very close to unstable regions and will ultimately escape from this resonance. By combining the inclination distributions of CSTs, CLTs and CTTs, resonance sweeping capture may exclude the existence of Twotinos with $i>50^{\circ}$. Moreover, this very mechanism predicts that the probability of discovering Twotinos with higher inclinations $i>25^{\circ}$ than observed values, near the ecliptic plane, is only $\sim0.1$ per cent.

Analytic arguments tell us that the ratio of CLTs to CTTs is largely determined by two factors: Neptune's migration time-scale and initial particle eccentricity (Murray-Clay \& Chiang 2005). In addition, we found that the influence of initial particle inclination $i_0$ could also be great on this ratio, provided small particle eccentricity of $e\sim0.01$. When $i_0\le10^{\circ}$, a particle is captured first into the symmetric 1:2 NMMR when Neptune migrates outward, then it will follow the dynamical pathway described in Murray-Clay \& Chiang (2005). Due to the long migration time-scale of $\tau=2\times10^7$ yr, such symmetric librator is preferentially caught into the leading rather than the trailing asymmetric resonance. However, for the case of $i_0\ge20^{\circ}$, a particle captured by the sweeping 1:2 NNMR from a nearly circular orbit would never firstly fall into the symmetric resonance. This is because the libration of $\sigma$ around $180^{\circ}$ is not permitted for orbits with $i>15^{\circ}$ and $e<e_c(i)$, as we have learned in the semi-analysis part. As a result, in the subsequent evolution, the probability of the high-inclination captive to enter the leading or trailing islands is not certain but would be continuously varying with $20^{\circ} \le i_0 \le 50^{\circ}$.

Among captured Twotinos from our simulations, we obtained a total of 4 potential $\omega$-librators: 1 in the symmetric resonance, 3 in the leading asymmetric resonance. These Kozai Twotinos all come from the cases of $i_0\le10^{\circ}$. Despite somewhat excitation due to the Kozai mechanism, their $i$ would always undergo low value episode below $15^{\circ}$ and end at $<12^{\circ}$. This outcome seems compatible with today's observed void of Kozai Twotinos residing in the trailing islands or having $i>15^{\circ}$. 

Nevertheless, the planet migration model suffers from two major problems: (1) the fraction of captured Twotinos with high inclinations is 1--2 orders of magnitude too low to account for present observations; (2) the capture efficiency of the Kozai mechanism that we obtain is only $\sim0.4$ per cent, whereas the observed value is $\sim8.3$ per cent. These discrepancies argue that an unknown fraction of currently observed Twotinos is not likely to originate by the resonance sweeping mechanism from a planetesimal disk interior to 47.8 au. These particular objects indicate that there is a distinct source region for the high-inclination Twotinos, and they might be able to promise new insight into the dynamical history of the Kuiper belt. We expect our future works could tell.

%______________________________________________________________

\section*{Acknowledgments}

This work was supported by the Natural Science Foundation of China (NSFC, Nos. 11003008, 11178006, 11333002), and the National `973' Project (No. 2013CB834103). LYZ has to thank the financial support of NSFC under grant No. 11073012  and the National  `973' Project (No. 2013CB834900). SYS has to acknowledge grant provided by the NSFC (No. 11078001). The authors would like to express their thanks to the anonymous referee for the valuable comments. JL is also grateful to Dr. Xiaosheng Wan for helpful discussions.

\end{document}